\documentclass[oneside]{elsart}
\usepackage{graphicx}
\usepackage[figuresright]{rotating}
\usepackage{amsmath}

\def\intablecenterline#1{\hfill#1\hfill} 
\def\und{} 
\def\ETTunderline#1{\ifmmode\let\und=\underline\else 
\let\und=\underbar\fi\und{#1}}

\def\mathline{\ifmmode \vert\else$\vert$\fi}

\def\ETTstretch#1{\message{[[Sorry, you can't use Stretch with ETT]]}} 
 
\def\ETTexpand#1{\message{[[Sorry, you can't use Expand with ETT]]}} 
 
\def\xbackslash{\ifmmode \backslash\else$\backslash$\fi}

\def\quoteswitch{} 
\def\leftquotemarks{\ifmmode{}{}''\else``%
\let\quoteswitch=\rightquotemarks\fi} 
\def\rightquotemarks{\ifmmode{}{}''\else"\let\quoteswitch=\leftquotemarks\fi} 
\let\quoteswitch=\leftquotemarks 
\def\p$&-\${\overline{p}} 
\def\ETTstack#1#2{\ifmmode\ifx\int#1\displaystyle#1\limits#2\else 
\ifx\xint#1\displaystyle\int\limits#2\else 
\ifx\sum#1\displaystyle#1\limits#2\else%
\ifx\xsum#1\displaystyle\sum\limits#2\else 
\mathop{#1}\limits#2\fi\fi\fi\fi\else#1\mathop{#2}\limits\fi} 
 
\def\ETThalign#1{{\let\centerline=\intablecenterline 
\ifmmode\vcenter{\everycr={\noalign{\vskip1.5pt}}%
\halign{$\strut##$&$\strut##$&$\strut##$&$\strut##$&$\strut##$&%
$\strut##$&$\strut##$&$\strut##$&$\strut##$&$\strut##$&$\strut##$&%
$\strut##$&$\strut##$&$\strut##$&$\strut##$&$\strut##$&$\strut##$&%
$\strut##$&$\strut##$&$\strut##$\cr#1\crcr}\everycr={}}\else%
\hbox{$\vcenter{\everycr={\noalign{\vskip1.5pt}} 
\halign{\strut##&\strut##&\strut##&\strut##&%
\strut##&\strut##&\strut##&\strut##&\strut##&\strut##&\strut##&%
\strut##&\strut##&\strut##&\strut##&\strut##&\strut##&\strut##&%
\strut##&\strut##\cr#1\crcr}\everycr={}}$}\fi}} 
 
\gdef\ETTcolumn#1{\ifmmode%
\hbox{$\vcenter{\everycr{\noalign{\vskip-2pt}}%
\let\centerline=\intablecenterline 
\halign{\strut\hfill$##$\hfill\cr 
#1\crcr}\vskip-2pt}$}\else 
\hbox{$\vcenter{\offinterlineskip%
\let\centerline=\intablecenterline 
\halign{\strut\hfill##\hfill\cr 
#1\crcr}}$}\fi} 
 
\def\ETTtable#1{\ifmmode\vcenter{\everycr={\noalign{\vskip1.5pt}}%
\let\centerline=\intablecenterline%
\halign{\hfil$\vcenter{\hbox{$\strut##$}}$\hfil\ &\ %
\hfil$\vcenter{\hbox{$\strut##$}}$\hfil\ &\ %
\hfil$\vcenter{\hbox{$\strut##$}}$\hfil\ &\ %
\hfil$\vcenter{\hbox{$\strut##$}}$\hfil\ &\ %
\hfil$\vcenter{\hbox{$\strut##$}}$\hfil\ &\ %
\hfil$\vcenter{\hbox{$\strut##$}}$\hfil\ &\ %
\hfil$\vcenter{\hbox{$\strut##$}}$\hfil\ &\ %
\hfil$\vcenter{\hbox{$\strut##$}}$\hfil\ &\ %
\hfil$\vcenter{\hbox{$\strut##$}}$\hfil\ \cr 
#1\crcr}\everycr={}}\vrule depth3pt width0pt\relax\else 
\hbox{$\vcenter{\everycr={\noalign{\vskip1.5pt}}%
\let\centerline=\intablecenterline%
\halign{\hfil$\vcenter{\hbox{\strut##}}$\hfil\ &\ %
\hfil$\vcenter{\hbox{\strut##}}$\hfil\ &\ %
\hfil$\vcenter{\hbox{\strut##}}$\hfil\ &\ %
\hfil$\vcenter{\hbox{\strut##}}$\hfil\ &\ %
\hfil$\vcenter{\hbox{\strut##}}$\hfil\ &\ %
\hfil$\vcenter{\hbox{\strut##}}$\hfil\ &\ %
\hfil$\vcenter{\hbox{\strut##}}$\hfil\ &\ %
\hfil$\vcenter{\hbox{\strut##}}$\hfil\ &\ %
\hfil$\vcenter{\hbox{\strut##}}$\hfil\ \cr 
#1\crcr}\everycr={}}$}\vrule depth3pt width0pt\relax\fi} 
 
\def\ETTast{{\raise.25ex\hbox{$\ast$}}} 
 
\newif\ifinmath 
\newbox\superbox 
\newbox\superboxtwo 
 
\def\changetomath{$} 
 
\def\ETTsuperimpose#1#2{\ifmmode\let\check=\changetomath%
\else\let\check=\relax\fi%
\setbox\superbox=\hbox{\check#1\check}%
\setbox\superboxtwo=\hbox{\check#2\check}%
\ifdim\wd\superbox>\wd\superboxtwo%
\copy\superbox\hskip-\wd\superbox\hbox to 
\wd\superbox{\hfill\check#2\check\hfill}%
\else%
\copy\superboxtwo\hskip-\wd\superboxtwo\hbox to 
\wd\superboxtwo{\hfill\check#1\check\hfill}\fi} 
 
\def\hb{{\hskip3pt}} 
\let~=\hb

\newcount\moveup \newcount\moveover 
 
\def\lookagain{\futurelet\next\parser} 
 
\def\parser#1{\def\endb{}\ifx\next /\let\go=\relax \else\ifx\next r 
\global\advance\moveover by22\else 
\ifx\next l\global\advance\moveover by-22 
\else \ifx\next u\global\advance\moveup by22 
\else \ifx\next d\global\advance\moveup by-22\fi\fi\fi\fi 
\let\go=\lookagain\fi\go} 
 
\def\ETTadjust#1#2{\moveover=0 \moveup=0\setbox0=\hbox{\lookagain#1/}%
\divide\moveover by10 \divide\moveup by10\setbox1=\hbox{#2}%
\vtop to0pt{\hbox to\wd1{\hskip\moveover pt\raise\moveup pt\hbox{#2}\hss}%
\vss}} 
 
\newcount\upcounter 
\newcount\overcounter 
 
\newif\ifaz \newif\ifrz \newif\iflz \newif\ifdz \newif\ifuz 
 
\def\nonglobalreset{\azfalse\lzfalse\uzfalse%
\dzfalse\rzfalse}%
\long\gdef\ETTbox #1 #2{\def\parse{#1}%
\nonglobalreset\expandafter\looker\parse{}{}{}\boxer{\vbox{%
\let\centerline=\intablecenterline\halign{%
\strut##\cr#2\crcr}}}} 
 
\long\def\boxer#1{\ifaz\rztrue\lztrue\uztrue\dztrue\fi%
\hbox{$\,\,\vcenter{\ifuz\hrule\fi\hbox{\iflz\vrule\fi%
\hskip2pt\vbox{\vskip2pt\vbox{#1}\vskip2pt}%
\hskip2pt\ifrz\vrule\fi}\ifdz\hrule\fi}\,\,$}} 
 
\def\looker #1#2#3#4{%
\ifx#1A\aztrue\else\ifx#2a\aztrue\else
\ifx#1R\rztrue\else 
\ifx#2R\rztrue\else 
\ifx#3R\rztrue\else 
\ifx#4R\rztrue\fi\fi\fi\fi\fi 
\ifx#1L\lztrue\else 
\ifx#2L\lztrue\else 
\ifx#3L\lztrue\else 
\ifx#4L\lztrue\fi\fi\fi\fi 
\ifx#1U\uztrue\else 
\ifx#2U\uztrue\else 
\ifx#3U\uztrue\else 
\ifx#4U\uztrue\fi\fi\fi\fi 
\ifx#1D\dztrue\else 
\ifx#2D\dztrue\else 
\ifx#3D\dztrue\else 
\ifx#4D\dztrue\fi\fi\fi\fi\fi} 
 
\def\tie{\ifmmode\sim\else 
\hbox{\vtop to0pt{\hbox{\lower5pt\hbox{\~{}}}\vss}}\fi} 
\def\<{\ifmmode <\else $<$\fi} 
\def\>{\ifmmode >\else $>$\fi} 
 
\def\caret{\ifmmode{\vtop to0pt{\hbox{\lower7pt\hbox{$\hat{\vphantom{.}}$}} 
\vss}}\else \^{}\fi} 
 
\def\lcurlybracket{\ifmmode\{\else $\{$\fi} 
\def\rcurlybracket{\ifmmode\}\else $\}$\fi} 
 
\newdimen\tempdimen 
 
\def\smdarkP{\ \hbox{\global\setbox0=\hbox{\smallbackp}\vrule height\ht0 
depth\dp0 width0pt\vrule\hskip.6pt\vrule}\tempdimen=\ht0%
\advance\tempdimen by-.4pt%
\hskip-1pt\raise1.4pt\hbox{$\scriptstyle\bullet$}%
\hskip-2pt\llap{\raise\tempdimen\hbox to3pt{\hrulefill}}\ } 
 
\def\xxdarkP{\ \hbox{\global\setbox0=\hbox{\P}\vrule height\ht0 
depth\dp0 width0pt\vrule\hskip.6pt\vrule}\tempdimen=\ht0%
\advance\tempdimen by-.4pt\hskip-1pt\raise3pt\hbox{$\displaystyle\bullet$}%
\hskip-2pt\llap{\raise\tempdimen\hbox to4pt{\hrulefill}}\ }

\def\smallbackp{{\smallsymbol\char'173}}

\def\return{\hbox{\ \unskip{$\bf\leftarrow$\hskip-.5pt\raise2.4pt%
\hbox{\vrule height 2.5pt}}}}

\def\xxTM{\raise1ex\hbox{\amrseven TM}} 
 
\def\TMfive{\raise1ex\hbox{\amrfive TM}}

\def\xxeqcirc{\buildrel \lower1.5pt\hbox{$\scriptstyle\circ$}\over =} 
\def\smeqcirc{\buildrel \lower1.5pt\hbox{$\scriptscriptstyle\circ$}\over{\scriptstyle =}}

\def\xxeqdkcirc{\buildrel \lower1.5pt\hbox{$\scriptscriptstyle\bullet$}\over =} 
\def\smeqdkcirc{\buildrel \lower1.5pt\hbox{$\scriptscriptstyle\bullet$}\over{\scriptstyle =}}

\def\xxtridots{\ \unskip\raise4pt\hbox 
to1em{\hfil.\hfil}\llap{\hbox to1em{\hfil.\ .\hfil}}} 
 
\def\smtridots{\ \unskip\raise3pt\hbox 
to1em{\hfil.\hfil}\llap{\hbox to1em{\hfil.$\,$.\hfil}}}

\def\sqr#1#2{{\vcenter{\hrule height.#2pt 
\hbox{\vrule width.#2pt height#1pt\kern#1pt 
\vrule width.#2pt} 
\hrule height.2pt}}}

\def\xint{{\mathchoice{\int}{\displaystyle\int}{\int}{\int}}} 
\def\xsum{{\mathchoice{\sum}{\displaystyle\sum}{\sum}{\sum}}} 
 
\def\GermanS{\ifmmode\hbox{\ss}\else\ss\fi} 
\def\primeaccent{\ifmmode\hbox{\rm\char19}\else{\rm\char19}\fi} 
\def\underaccent{\ifmmode\hbox{\rm\char24}\else{\rm\char24}\fi} 
\def\EnglishPound{\ifmmode\hbox{\it\$}\,\else{\it\$}\fi} 
 
\newcommand{\be}{\begin{equation}}
\newcommand{\ee}{\end{equation}}
\newcommand{\bea}{\begin{eqnarray}}
\newcommand{\eea}{\end{eqnarray}}
\newcommand{\beas}{\begin{eqnarray*}}
\newcommand{\eeas}{\end{eqnarray*}}
\newcommand{\bi}{\begin{itemize}}
\newcommand{\ei}{\end{itemize}}
\newcommand{\bn}{\begin{enumerate}}
\newcommand{\en}{\end{enumerate}}

\def\lambar{{ \lambda \mkern-10mu\raise.5ex\hbox{--} }}

\def\thus{{ .. \mkern-7.5mu\raise.9ex\hbox{.} }\  }

\def\ba2#1#2{${\overline{#1}}^{#2}$}
\def\anti#1#2{\vbox{\ialign{##\crcr
     \hrulefill$\smash{\phantom{\scriptstyle#2}}$\crcr
     \noalign{\kern-1pt\nointerlineskip\vskip 0.25ex}
     $\hfil{#1}^{#2}\hfil$\crcr}}}
\def\anth#1#2{\vbox{\ialign{##\crcr
    \hrulefill$\smash{\phantom{\scriptstyle#2}}$\crcr
    \noalign{\kern-0.5pt\nointerlineskip\vskip 0.25ex}
    $\hfil{#1}^{#2}\hfil$\crcr}}}
\def\rless{{ r \mkern-.5mu\raise-.2ex\hbox{\tiny{<}} }}
\def\rmore{{ r \mkern-.5mu\raise-.2ex\hbox{\tiny{>}} }}

\newcommand{\Ket}{{\rm\bf\goodfontB Ket}}

\newcommand{\Bra}{{\rm\bf\goodfontB Bra}}

\newcommand{\had}{{\rm\bf\goodfontB had}}

\newcommand{\Had}{{\rm\bf\goodfontB Had}}

\newcommand{\RotX}{{\rm\bf\goodfontB RotX}}
\newcommand{\RotY}{{\rm\bf\goodfontB RotY}}
\newcommand{\RotZ}{{\rm\bf\goodfontB RotZ}}

\newcommand{\Purity}{{\rm\bf\goodfontB Purity}}
\newcommand{\Entropy}{{\rm\bf\goodfontB Entropy}}
\newcommand{\Fidelity}{{\rm\bf\goodfontB Fidelity}}

\newcommand{\HALL}{{\rm\bf\goodfontB HALL}}

\newcommand{\SP}{{\rm\bf\goodfontB SP}}

\newcommand{\cnot}{{\rm\bf\goodfontB CNOT }}
\newcommand{\NOT}{{\rm\bf\goodfontB NOT }}

\newcommand{\cphase}{{\rm\bf\goodfontB CPHASE }}
\newcommand{\crot}{{\rm\bf\goodfontB CROT }}

\newcommand{\QDENS}{{\rm\bf\goodfontB QDENSITY\,}}

\newcommand{\PTr}{{\rm\bf\goodfontB PTr }}

\newcommand{\Toffoli}{{\rm\bf \goodfontB Toffoli }}

\newcommand{\Swap}{{\rm\bf \goodfontB Swap }}

\newcommand{\ControlledY}{{\rm\bf \goodfontB ControlledY }}
\newcommand{\ControlledX}{{\rm\bf \goodfontB ControlledX }}

\def\goodfontB{\usefont{T1}{phv}{n}{n}\fontsize{9pt}{9pt}\selectfont}

\newcommand{\cal}{\mathcal}

\input{Qcircuit}

\renewcommand{\Qcircuit}[1][0em]{\xymatrix @*[o] @*=<#1>}

\begin{document}
\begin{frontmatter}
\title{QDENSITY - A MATHEMATICA  QUANTUM COMPUTER SIMULATION}

\author[pit,cea]{Bruno Juli\'a-D\'{\i}az},
\author[pit,NCS]{Joseph M. Burdis}
\and
\author[pit]{Frank Tabakin}

\address[pit]{Department of Physics and Astronomy \\
University of Pittsburgh\\
Pittsburgh, PA, 15260}

\address[cea]{DAPNIA, DSM,
CEA/Saclay\\
91191 Gif-sur-Yvette, France}

\address[NCS]{Department of Mathematics\\
North Carolina State University\\
Raleigh, NC 27695}

\begin{abstract}
This Mathematica 5.2 package~\footnote{QDENSITY is available at
http://www.pitt.edu/\~\,\!tabakin/QDENSITY} is a simulation of a
Quantum Computer. The program provides a modular, instructive
approach for generating the basic elements that make up a quantum
circuit. The main emphasis is  on using the density matrix, although
an approach using state vectors is also implemented in the package.
The package commands are defined in {\it Qdensity.m} which contains
the tools needed in quantum circuits, e.g. multiqubit kets,
projectors, gates, etc. Selected examples of the basic commands are
presented here and a tutorial notebook, {\it Tutorial.nb} is
provided with the package (available on our website) that serves as a full guide to the package.
Finally, application is made to a variety of relevant cases,
including Teleportation, Quantum Fourier transform, Grover's search
and Shor's algorithm, in separate notebooks: {\it QFT.nb}, {\it
Teleportation.nb}, {\it Grover.nb} and {\it Shor.nb} where each
algorithm is explained in detail. Finally, two examples of the
construction and manipulation of cluster states, which are part of
``one way computing" ideas, are included as an additional tool in
the notebook {\it Cluster.nb}. A Mathematica palette containing most
commands in QDENSITY is also included: {\it QDENSpalette.nb} .
\vspace{1pc}
\end{abstract}
\end{frontmatter}

\newpage
\noindent{\bf Program Summary}

{\it Title of program:} QDENSITY\\
{\it Catalogue identifier:}\\
{\it Program summary URL:} http://cpc.cs.qub.ac.uk/summaries\\
{\it Program available from:} CPC Program Library, Queen's University of Belfast, N. Ireland\\
{\it Operating systems:} Any which supports Mathematica; tested under Microsoft Windows XP, Macintosh OS X, and Linux FC4.\\
{\it Programming language used:} Mathematica 5.2\\
{\it Number of bytes in distributed program, including test code and 
documentation: 1185886}\\
{\it Distribution format:} tar.gz\\
{\it Nature of Problem:} Analysis and design of quantum circuits, 
quantum algorithms and quantum clusters.\\
{\it Method of Solution:} A Mathematica package is provided which contains commands to create and analyze quantum circuits. Several Mathematica notebooks containing relevant examples: Teleportation, Shor's Algorithm and Grover's search are explained in detail. A tutorial, Tutorial.nb is also enclosed. \\

\newpage
\tableofcontents

\section{INTRODUCTION}

There is already a rich Quantum Computing (QC) literature~\cite{Nielsen} which
holds forth the promise of using quantum interference and superposition
to solve otherwise intractable problems. The field has reached
the point that experimental realizations are of paramount importance
and theoretical tools towards that goal are needed: to gauge the
efficacy of various approaches, to understand the construction
and efficiency of the basic logical gates, and to delineate and
control environmental decoherence effects.

In this paper, a Mathematica~\cite{Mathematica} package provides
a simulation of a Quantum Computer that is both flexible and an
improvement over earlier such works~\cite{prevQC}. It is a bona
fide simulation in that its success depends on quantum interference
and superposition and is not just a simulation of the QC experience.
The flexibility is generated by a modular approach to all of the
initializations, operators, gates, and measurements, which then
can be readily used to describe the basic QC Teleportation~\cite{Teleportation}, Grover's
search~\cite{Grover,Groverslit} and Shor's factoring~\cite{Shor} algorithms. We also adopt a density
matrix approach as an organizational framework for introducing
fundamental Quantum Computing concepts in a manner that allows for
more general treatments, such as handling the dynamics stipulated by
realistic Hamiltonians and including environmental effects. That
approach allows us to invoke many of the dynamical theories based on
the time evolution of the density matrix. Since much of the code uses
the density matrix, we call it ``\QDENS," which stands for Quantum
computing with a density matrix framework. However, the code also
provides the tools to work directly with multi-qubit states as an
alternative to the density matrix description.

In section~\ref{sec2}, we introduce one qubit state vectors and associated
 spin operators, including rotations, and introduce commands from \QDENS.
The basic characteristics of the density matrix are then discussed in a
pedagogic manner
in section~\ref{sec3}. Then in  section~\ref{sec4}, methods for handling
multi-qubit operators, density matrices, and state vectors with commands from \QDENS are
presented. Also in that section, we show how to take traces, subtraces
and how to combine subtraces with general projection operators to
simulate projective measurements in a multi-qubit context. The basic  one, two
and three qubit gates (Hadamard, CNOT, CPHASE, Toffoli, etc.) needed for the QC circuits
are shown in section~\ref{sec5}.
The production of entangled states, such as the two-qubit Bell~\cite{Bell}
states, the three-qubit GHZ~\cite{GHZ} states, among others,~\cite{Werner} are
 illustrated in both density matrix and state vector renditions in  section~\ref{sec6}.
 In sections~\ref{sec7}-\ref{sec9},  Teleportation, Grover's search, and Shor's
 factoring algorithms are outlined,  with the detailed instructions relegated to
  associated notebooks.
Sample application to the cluster or ``one-way computing" model of QC is
presented in section~\ref{sec10}.  Possible future applications of \QDENS
are given in the conclusion
section~\ref{sec11}.

The basic logical gates used in the circuit model of Quantum Computing
are presented in a way that allows ease of use and hence permits one
to construct the unitary operators corresponding to well-know quantum
algorithms. These algorithms are developed explicitly in the Mathematica
notebooks as a demonstration of the application of \QDENS.
 A tutorial notebook (Tutorial.nb)  available
 on our web site guides the user through the requisite manipulations. Many examples
from \QDENS, as defined in the package file Qdensity.m, are
discussed throughout the text, which hopefully, with the tutorial
notebook, will help the user to employ this tool.

All these examples are instructive in two ways. One way is to
learn how to handle \QDENS for other future applications and
generalizations, such as studying entanglement measures, examining
the time evolution generated by experiment-based realistic
Hamiltonians, error correction methods, and the role of the environment
and its affect on coherence. Thus the main motivation for
emphasizing a density matrix formulation is that the time evolution
can be described, including effects of an environment, starting
from realistic Hamiltonians~\cite{Lindblad,Preskill}. Therefore, \QDENS
provides an introduction to methods that can be generalized to an
increasingly realistic description of a real quantum computer.

Another instructive feature is to gain insight into how quantum
superposition and interference are used in QC to pose and to answer
questions that would be inaccessible using a classical computer.
Thus we can form an initial description of a quantum multi-qubit
state and have it evolve by the action of carefully designed
unitary operators. In that development, the prime characteristics of
superposition and interference of probability amplitudes is cleverly
applied in Quantum Computing to enhance the probability of getting the
right answer to problems that would otherwise take an immense time
to solve.

In addition to applying \QDENS to the usual quantum circuit model for
QC, we have adapted it to examine the construction of cluster states
and the steps needed to reproduce general one qubit and two qubit
operations.  These cluster model examples are included to open the
door for future studies of the quite promising cluster model~\cite{Cluster} or
``one-way computing" approach for QC.

We sought to simulate as large a system of qubits as possible, using
new features of Mathematica.  Of course,  this code is a simulation
of a quantum computer based on Mathematica code run on a classical
computer. So it is natural that the simulation saturates memory for
large qubit spaces; after all, if the QC algorithms always worked
efficiently on a classical computer there would be no need for a
quantum computer.

{\bf Throughout the text, sample \QDENS commands are presented in
sections called ``Usage." The reader should consult Tutorial.nb for
more detailed guidance.}

\section{ONE QUBIT SYSTEMS}
\label{sec2}

The state of a quantum system is described by a wave function which
in general depends on the space or momentum coordinates of the
particles and on time. In Dirac's representation independent
notation, the state of a system is a vector in an abstract Hilbert
space $\mid \Psi(t)>$, which depends on time, but in that form one
makes no choice between the coordinate or momentum space
representation. The transformation between the space and momentum
representation is contained in a transformation bracket. The two
representations are related by Fourier transformation, which is the
way Quantum Mechanics builds localized wave packets. In this way,
uncertainty principle limitations on our ability to measure
coordinates and momenta simultaneously with arbitrary precision are
embedded into Quantum Mechanics (QM). This fact leads to operators,
commutators, expectation values and, in the special cases when a
physical attribute can be precisely determined, eigenvalue equations
with Hermitian operators. That is the content of many quantum texts.
Our purpose is now to see how to define a density matrix, to
describe systems with two degrees of freedom as needed for quantum
computing.

Spin, which is the most basic two-valued quantum attribute, is
missing from a spatial description. This subtle degree of freedom,
whose existence is deduced by analysis of the Stern-Gerlach
experiment, is an additional Hilbert space vector feature. For
example, for a single spin 1/2 system the wave function including
both space and spin aspects is: \be \Psi(\vec{r}_1,  t) \mid s\
m_s>, \ee where $\mid s \ m_s>$ denotes a state that is
simultaneously an eigenstate of the particle's total spin operator
$s^2 = s_x^2 +s_y^2+s_z^2$, and of its spin component operator
$s_z$. That is \be s^2 \mid s m_s> = \hbar^2 s (s+1) \mid s m_s>
\qquad s_z \mid s m_s> = \hbar m_s \mid s m_s> \,. \ee For a spin
1/2 system, we denote the spin up state as $\mid s m_s>\rightarrow
\mid \frac{1}{2},\frac{1}{2}> \equiv \mid 0>$, and the spin down
state as $\mid s m_s>\rightarrow \mid \frac{1}{2},-\frac{1}{2}>
\equiv \mid 1>$.

We now arrive at the definition of a one qubit state as a
superposition of the two states associated with the above $0$ and
$1$ bits: \be \mid \Psi> =a  \mid 0>+  b \mid 1>, \ee where $a
\equiv <0\mid \Psi>$ and $b\equiv <1\mid \Psi>$ are complex
probability amplitudes for finding the particle with spin up or
spin down, respectively. The normalization of the state
$<\Psi\mid\Psi> =1$, yields $\mid a \mid^2 + \mid b \mid^2=1$.
Note that the spatial aspects of the wave function are being
suppressed; which corresponds to the particles being in a fixed
location, such as at quantum dots.~\footnote{When these separated
systems interact, one might need to restore the spatial aspects of
the full wave function.}

An essential point is that a QM system can exist in a
superposition of these two bits; hence, the state is called a
quantum-bit or ``qubit." Although our discussion uses the notation
of a system with spin, it should be noted that the same discussion
applies to any two distinct states that can be associated with
$\mid 0> $ and $ \mid 1>$.  Indeed, the following section on the
Pauli spin operators is really a description of any system that
has two recognizable states.

\subsubsection{Usage }

  \QDENS includes commands for  qubit states as ket and
bra vectors. For example, commands \Ket[0], \Bra[0],\Ket[1],
\Bra[1], yield

\bea
{\rm In[1]} &:=&{\rm \bf Ket[0]} \nonumber \\
{\rm Out[1]}&:=& \left(
\begin{array}{l}
 1 \\
 0
\end{array}
\right) \nonumber
\eea
\bea
{\rm In[2]} &:=&{\rm \bf Ket[1]} \nonumber \\
{\rm Out[2]}&:=&{\bf \left(
\begin{array}{l}
 0 \\
 1
\end{array}
\right)} \nonumber  \eea

\bea
{\rm In[3]} &:=&{\rm \bf Bra[0]} \nonumber \\
{\rm Out[3]}&:=&{\bf ( 1\;\; 0 )} \nonumber  \eea

\bea
{\rm In[4]} &:=&{\rm \bf Bra[1]} \nonumber \\
{\rm Out[4]}&:=&{\bf ( 0\;\; 1 )} \nonumber  \eea

These are the computational basis states, i.e. eigenstates of the
spin operator in the z-direction.

States that are eigenstates of the spin operator in the x-direction
are invoked by the commands

\bea
{\rm In[5]} &:=&{\rm \bf BraX[0]} \nonumber \\
{\rm Out[5]}&:=&{\bf \Big( \frac{1}{\sqrt{2}} \;\;
\frac{1}{\sqrt{2}} \Big)} \nonumber  \eea

which is equivalent to:

\bea
{\rm In[6]}&:=&{\rm \bf (Bra[0]+Bra[1])}/\sqrt{2} \nonumber \\
{\rm Out[6]}&:=&{\bf \Big( \frac{1}{\sqrt{2}} \;\;
\frac{1}{\sqrt{2}} \Big)} \nonumber  \eea

Eigenstates of the spin operator in the y-direction are invoked
similarly the commands
${\rm \bf BraY[0]}$, ${\rm \bf BraY[1]}$, etc.

\subsection{The Pauli Spin Operator}

We use the case of a spin 1/2 particle to describe a quantum system
with two discrete levels; the same description can be applied to any
QM system with two distinct levels. The spin $\vec{s}$ operator is
related to the three Pauli spin operators $\sigma_x, \sigma_y,
\sigma_z$ by \be \vec{s} \equiv (\frac{\hbar}{2}) \vec{\sigma} , \ee
from which we see that $\vec{\sigma}$ is an operator that describes
the spin 1/2 system in units of $\frac{\hbar}{2}$. Since spin is an
observable, it is represented by a Hermitian operator,
$\vec{\sigma}^\dagger = \vec{\sigma}$. We also know that measurement
of spin is subject to the uncertainty principle, which is reflected
in the non-commuting operator properties of spin and hence of the
Pauli operators. For example, from the standard commutator property
for any spin $[s_x, s_y] = i \hbar s_z,$ one deduces that the Pauli
operators do not commute \be [  \sigma_x , \sigma_y ] =  2  i
\sigma_z \ . \ee\ This holds for all cyclic components so we have a
general form~\footnote{Here the Levi-Civita symbol is nonzero only
for cyclic order of components $ { i j k }= xyz, yzx ,zxy ,$ for
which $ \epsilon_{ i j k}=1.$ For anti-cyclic order of components ${
i j k}= xzy, zyx, yxz$ $\epsilon_{ ijk}=-1.$  It is otherwise zero.}
\be [ \sigma_i, \sigma_j ] = 2  i \epsilon_{ i j k}  \sigma_k \ .
\ee

An important property of the spin algebra is that the total
spin commutes with any component of the spin operator
$[s^2, s_i]=0$ for all $i$. The physical consequence is
that one can simultaneously measure the total spin of the
system and one component (usually $s_z$) of the spin. Only one
component is a candidate for simultaneous measurement because
the commutator $\nobreak{[s_x, s_y] = i \hbar s_z}$ is already an
uncertainty principle constraint on the other components. As
a result of the ability to measure $s^2$ and $s_z$ simultaneously,
the allowed states of the spin 1/2 system are restricted to being
spin-up and spin-down with respect to a specified fixed direction
$\hat{z}$, called the axis of quantization. States defined relative
to that fixed axis are called  ``computational basis" states, in
the QC jargon. The designation arises because as noted already
one can identify spin-up with a state $\mid 0>$, which designates
the digit or bit $0,$ and a spin-down state as $\mid 1>,$ which
designates the digit or bit $1$.

The fact that there are just two states (up and down) also implies
properties of the Pauli operators. We construct~\footnote{With the
definition $s_\pm\equiv s_x\pm  i s_y,$ and using the original
spin commutation rules, it follows that $[ s_\pm , s_z] = \mp
s_\pm ,$ which reveals that $s_\pm$ and hence also $\sigma_\pm$
are raising and lowering operators.  The general result,
including the limit on the total spin is $s_\pm \mid s\ m_s> =
\sqrt{s(s+1) -m_s(m_s\pm1)} \mid s\   m_s \pm 1>.$ } the raising
and lowering operators $\nobreak{\sigma_\pm = \sigma_x \pm i
\sigma_y}$, and note that the raising and lower process is bounded
\be
\sigma_+ \mid 0 > = 0 \qquad\sigma_{-} \mid 1 > = 0.
\ee
Hence, raising a two-valued state up twice or lowering it
twice yields a null (unphysical) Hilbert space; this property
tells us additional aspects of the Pauli operator. Since
\be
\sigma_\pm \sigma_\pm =( \sigma_x \pm i \sigma_y)^2 = \sigma_x^2-
\sigma_y^2 \pm ( \sigma_x  \sigma_y + \sigma_y  \sigma_x) =0,
\ee
we deduce that $ \sigma_x^2=\sigma_y^2$, and that the anti-commutator
\be
\{\sigma_x , \sigma_y\} \equiv \sigma_x  \sigma_y + \sigma_y  \sigma_x =0.
\ee
The anti-commutation property is thus a direct consequence of
the restriction to two levels.

The spin 1/2 property is often expressed as:
$ s^2 \mid s m_s > = \hbar^2 s ( s+1)  \mid s m_s >
= \frac{3}{4} \hbar^2  \mid s m_s >
= \frac{\hbar}{4}^2 \sigma^2  \mid s m_s >.$
We have
$\sigma^2 =3 = \sigma_x^2 +  \sigma_y^2 + \sigma_z^2 =2 \sigma_x^2 +1,$
where we use the above equality $\sigma_x^2=\sigma_y^2,$ and
from the $\hat{z}$ eigenvalue equation the property
$\sigma_z^2 = 1,$ to deduce that
\be
\sigma_x^2 = \sigma_y^2 =\sigma_z^2=1.
\ee

Another useful property of the Pauli matrices is obtained by
combining the above properties and commutator and anti-commutator
into one expression for a given spin 1/2 system
 \be \sigma_i
\sigma_j = \delta_{i j} + i \epsilon_ {i j k }  \sigma_k, \ee
where
indices $ i ,j, k $ take on the values $x,y,z,$ and repeated indices
are assumed to be summed over. For two general vectors, this becomes
 \be
 (\vec{\sigma}\cdot \vec{A} )  (\vec{\sigma}\cdot \vec{B} ) = \vec{A}
  \cdot \vec{B} + i (\vec{A} \times \vec{B}) \cdot \vec{\sigma}.
 \ee
For $\vec{A}=\vec{B}= \vec{\eta},$ a unit vector
$(\vec{\sigma}\cdot \vec{\eta} )^2 =1$, which will be useful later.

These operator properties can also be represented by the
Pauli-spin matrices, where we identify the matrix elements by
\be
< s\ m'_s \mid \sigma_z \mid s\ m_s > \longrightarrow
 \left(
\begin{array}{lc}
1 & 0\\
0 & -1
\end{array}\right)  \, .
\ee

Similarly for the $x-$ and $y-$component spin operators
\be
< sm'_s \mid \sigma_x \mid s\ m_s > \longrightarrow
\left(
\begin{array}{lccr}
0&&& 1\\
1&&& 0
\end{array}\right)
\qquad
< sm'_s \mid \sigma_y \mid sm_s > \longrightarrow \left(
\begin{array}{lcr}
0 && -i \\
i && 0
\end{array}\right) \, .
\ee  These are all Hermitian matrices $\sigma_i =
\sigma^\dagger_i$.

Also, the matrix realization for the raising and lowering
operators are:
 \be \sigma_{+} = \left(
\begin{array}{lccr}
0 &&& 2 \\
0 &&& 0
\end{array}\right)
\hspace{.25in} \sigma_{-}= \left(
\begin{array}{lccr}
0 &&& 0 \\
2 &&& 0
\end{array}\right) \, .
\ee
Here $\sigma_+^\dagger =\sigma_{-}.$

Note that these $2 \times 2$  Pauli matrices are traceless
${\rm Tr} [ \vec{\sigma}] = 0, $  unimodular $\sigma^2_i = 1$
and have unit determinant $\mid \det \sigma_i \mid = 1$. Along
with the unit operator
\be \sigma_0= {\bf 1}  \equiv
\left(
\begin{array}{lccr}
1 &&& 0 \\
0 &&& 1
\end{array}\right) ,
\ee
the four Pauli operators form a basis for the expansion of
any spin operator in the single qubit space. For example, we
can express the rotation of a spin as such an expansion and
later we shall introduce a density matrix for a single qubit
in the form
$\rho = a + \vec{b} \cdot \vec{\sigma} = a + b\  \vec{n} \cdot \vec{\sigma}$
to describe an ensemble of particle spin directions as occurs
in a beam of spin-1/2 particles.

In \QDENS, we denote the four matrices by $\sigma_i$ where $i=0$ is
the unit matrix and $i=1,2,3$ corresponds to the components $x,y,$
and $z$. To produce the Pauli spin operators in \QDENS, one can use
either the Greek form or the expression $s[i]$.

\subsubsection{Usage }

  \QDENS includes commands for  the Pauli operators.
   For example, there are three equivalent ways to invoke the
   Pauli $\sigma_y$ matrix in \QDENS:

\bea
{\rm In[7]} &:=&{\rm \bf \sigma_y} \nonumber\\
{\rm Out[7]}&:=&{\bf  \left(
\begin{array}{ll}
 0 & -i \\
 i & 0
\end{array}
\right)} \nonumber  \eea

\bea
{\rm In[8]} &:=&{\rm \bf s[2]} \nonumber\\
{\rm Out[8]}&:=& {\bf \left(
\begin{array}{ll}
 0 & -i \\
 i & 0
\end{array}
\right)} \nonumber  \eea  The third way is to use the commands {\bf
Sigma0}, {\bf Sigma1},{\bf Sigma2}, or {\bf Sigma3}.

Note that
\bea
{\rm In[9] }&:=& {\rm \bf \sigma _2}\,.\,{\rm \bf KetY[0]-KetY[0] } \nonumber\\
{\rm Out[9]}&:=& {\bf \left(
\begin{array}{l}
 0  \\
 0
\end{array}
\right)} \nonumber  \eea

and

\bea
{\rm In[10] }&:=& {\rm \bf \sigma _2}\,.\,{\rm \bf KetY[1]+KetY[1] } \nonumber\\
{\rm Out[10]}&:=& {\bf \left(
\begin{array}{l}
 0  \\
 0
\end{array}
\right)} \nonumber \eea

confirm that {\rm \bf KetY[0]} and {\rm \bf KetY[1]} are indeed  eigenstates of
$\sigma_y.$  Note that the $ \cdot $ is used to take the dot
product.

\subsection{Pauli Operators in Hilbert Space}

It is often convenient to express the above matrix properties
in the form of operators in Hilbert space. For a general
operator $\Omega$, using closure, we have
\be
\Omega = \sum_n \sum_{n'} \mid n> < n \mid \Omega \mid n'> <n'\mid \, .
\ee
For the four Pauli operators this yields:
\bea
\sigma_0 &=& \mid 0 > < 0 \mid +   \mid 1 > < 1 \mid  \nonumber \\
\sigma_1 &=& \mid 0 > < 1 \mid +   \mid 1 > < 0 \mid  \nonumber \\
\sigma_2 &=& - i \mid 0 > < 1 \mid +   i \mid 1 > < 0 \mid  \nonumber \\
\sigma_z &=& \mid 0 > < 0 \mid -    \mid 1 > < 1 \mid  \, .
\eea
Taking matrix elements of these operators reproduces the
above Pauli matrices. Also note we have the general trace property
\be
{\rm Tr} \mid a><b\mid =  \sum_n < n \mid a >< b \mid n>
=  \sum_n < b \mid n> < n \mid a > =
 < b \mid a>,
\ee where $\mid n>$ is a complete orthonormal (CON)
basis~\footnote{For a CON basis, we have closure $ \sum_n  \mid
n>\ < n \mid= {\bf 1}.$}. Applying this trace rule to the above
operator expressions confirms that ${\rm Tr}[\sigma_x] ={\rm
Tr}[\sigma_y]={\rm Tr}[\sigma_z]=0,$ and ${\rm Tr}[\sigma_0]=2$.

Another useful trace rule is
${\rm Tr}[\; \Omega \ \mid a><b \mid \, ] = <b \mid \Omega \mid a>$.

\subsection{Rotation of Spin}

\begin{figure}[h]
\begin{center}
\includegraphics[width=8pc]{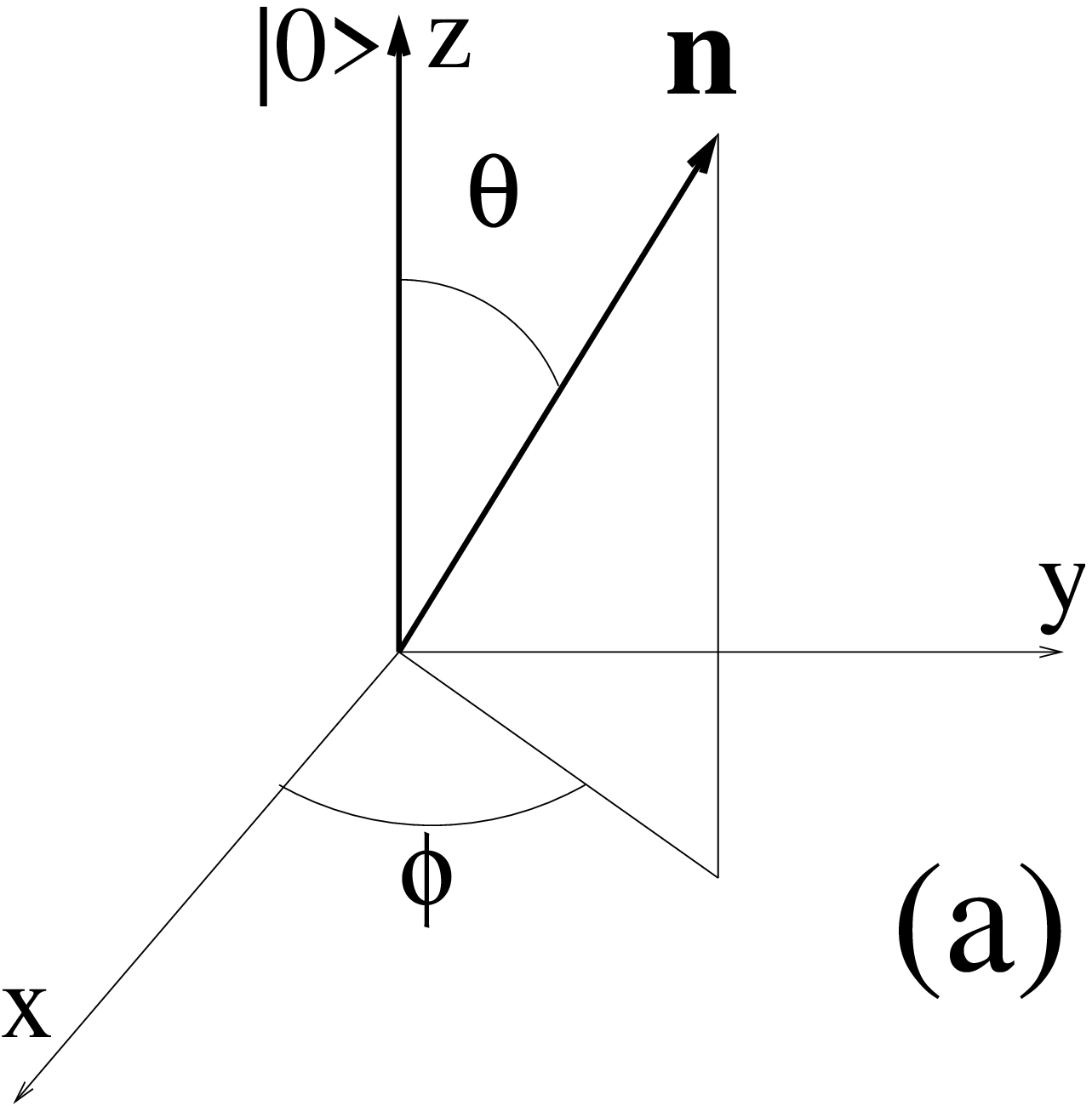}
\hspace*{20pt}
\includegraphics[width=8pc]{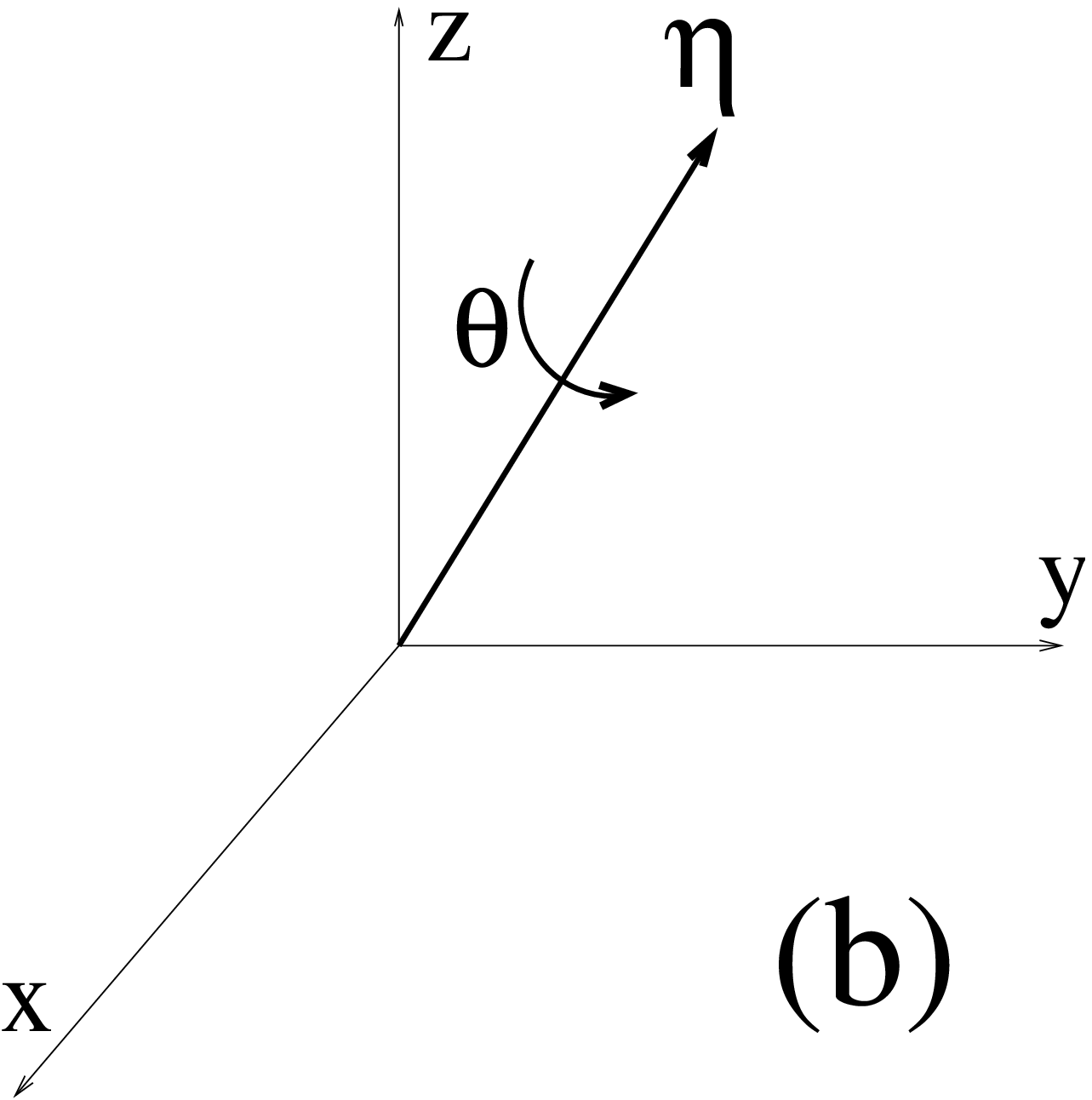}
\end{center}
\caption{Active rotation {\it to} a direction $\hat{n}$  (a); and active rotation
{\it around} a vector $\hat{\eta}$ (b). }
\protect\label{spinrot}
\end{figure}

Another way to view the above superposition, or qubit, state is
that a state originally in the $\hat{z}$ direction $\mid 0>$, has
been rotated {\it to a new direction} specified by a unit vector
$\hat{n}= ( n_x,n_y, n_z)=( \sin \theta \cos \phi, \sin\theta\sin
\phi, \cos \theta),$ as shown in Fig.~\ref{spinrot}.

The rotated spin state
\be
\mid \hat{n}> =
\cos(\theta/2) e^{-i \phi/2} \mid 0> +\sin(\theta/2) e^{+i \phi/2} \mid 1>
= \left(
\begin{array}{l}
\cos(\theta/2) e^{-i \phi/2}  \\
\sin(\theta/2) e^{+i \phi/2}
\end{array}\right) \, ,
\ee
is a normalized eigenstate of the operator
\be
\vec{\sigma}\cdot \hat{n} =
\left(
\begin{array}{lc}
n_z & n_x-i n_y \\
n_x+i n_y & -n_z
\end{array}\right)
= \left(
\begin{array}{lc}
\cos\theta & \sin \theta e^{-i \phi} \\
\sin \theta e^{i \phi} & -\cos\theta
\end{array}\right) \,.
\ee
We see that
\be
\vec{\sigma}\cdot \hat{n} \mid \hat{n}> =  \mid
\hat{n}>.
\ee
The half angles above reflect the so-called spinor nature
of the QM state of a spin 1/2 particle in that a double full
rotation is needed to bring the wave function back to its
original form.

These rotated spin states allow us to pick a different axis
of quantization $\hat{n}$ for each particle in an ensemble
of spin 1/2 systems. These states are normalized
$<\hat{n} \mid \hat{n}>=1,$ but are not orthogonal
$<\hat{n}' \mid \hat{n}>\neq 1$, when the $\hat{n}$ angles
$\theta,\phi$ do not equal the $\hat{n}'$ angles $\theta',\phi'.$

Special cases of the above states with spin pointing in the
directions $\pm\hat{x}$ and $\pm\hat{y}$ are within phases:  \bea \mid \pm x>
&=&\frac{1}{\sqrt{2}} \left(
 \begin{array}{c}
\ 1  \\
\pm 1
\end{array}\right) \rightarrow \frac{\mid 0 > \pm \mid 1> }{\sqrt{2}} \, ,
\nonumber \\
\mid \pm y> &=& \frac{1}{\sqrt{2}}\left(
\begin{array}{c}
\ 1  \\
\pm i
\end{array}\right)\rightarrow \frac{\mid 0 > \pm  i \mid 1> }{\sqrt{2}} \,.
\eea  Hilbert space versions are also shown above.

Rotation can also be expressed as a rotation of an initial spin-up
system {\it about a rotation axis} $\hat{\eta}$ by an angle
$\gamma$. Thus an operator $R_\gamma \equiv e^{-i
\frac{\gamma}{2}\vec{ \sigma} \cdot\hat{\eta}}, $ acting as \be
\mid \Psi > =R_\gamma \mid 0>  \ee can also rotate the spin system
state to new direction. This rotation operator can be expanded in
the form \be R_\gamma = e^{-i \frac{\gamma}{2}\vec{ \sigma}
\cdot\hat{\eta}}= \cos \frac{\gamma}{2}\  \sigma_0 -i \sin
\frac{\gamma}{2}\  \vec{\sigma}\cdot \hat{\eta}, \label{eq:rot}
\ee which follows from the property that $(\vec{\sigma}\cdot
\hat{\eta})^2 = \hat{\eta} \cdot \hat{\eta}+ i (\hat{\eta}\times
\hat{\eta}) \cdot \vec{\sigma} =1 \, .$ A special case of the
state generated by this rotation $R_\gamma \mid 0 > $  is a
$\gamma=\pi/2$ rotation about the  $ \hat{\eta}\rightarrow
\hat{y}$ axis.  Then the rotation operator is \be R_{\pi/2} =
e^{-i \frac{\pi}{4} \sigma_y} = \cos \frac{\pi}{4} \sigma_0 - i
\sin \frac{\pi}{4}\ \sigma_y \ . \ee Introducing the Pauli
matrices, this becomes \be R_{\pi/2} = \frac{1}{\sqrt{2}} \left(
\begin{array}{ccc}
  1 && -1 \\
  1 && \ \ 1
\end{array}\right) .
\ee This rotation about an axis again yields the
same result; namely, \be R_{\pi/2}  \mid 0 >= \frac{1}{\sqrt{2}}
\left(
\begin{array}{ccl}
1 && -1 \\
1&& \  \ 1
\end{array}\right)  \cdot
\frac{1}{\sqrt{2}}  \left(
\begin{array}{l}
  1  \\
  0
\end{array}\right) =\frac{1}{\sqrt{2}}  \left(
\begin{array}{l}
  1  \\
  1
\end{array}\right)  \   .
\ee
Similar steps apply for a rotation about the $\hat{x}$ axis by
$\gamma=\pi/2,$ which yields the earlier $\mid \pm y >$ states.

From normalization of the rotated state
$<\Psi\mid\Psi>=\nobreak{<0\mid R^\dagger_{\gamma}R_{\gamma}\mid
0>}= \nobreak{<0\mid 0>}=1,$ we see that the rotation is a unitary
$R^\dagger_{\gamma} R_{\gamma} ={\bf 1}$ operator.

\subsubsection{Usage}

The trace of the Pauli operators is invoked in \QDENS by:\\

\bea
{\rm In[1] }&:=& {\rm \bf Tr[\sigma _2]}\nonumber\\
{\rm Out[1]}&:=& {\bf 0}  \nonumber \eea

Rotation about the X axis is represented by \\

\bea
{\rm  In[2] }&:=& {\rm \bf RotX[\theta]}\nonumber \\
{\rm  Out[2]}&:=& {\bf \left(
\begin{array}{ll}
 \text{Cos}\left[\frac{\theta}{2}\right] & -i \text{Sin}\left[\frac{\theta}{2}\right] \\
 -i \text{Sin}\left[\frac{\theta}{2}\right] & \text{Cos}\left[\frac{\theta}{2}\right]
\end{array}\right)} \nonumber
\eea

Commands for other directions are described in {\it  Tutorial.nb};
see RotX[$\theta$], RotY[$\theta$], Rotqbit[vec,$\theta$].

\subsection{One Qubit Projection}

For a one qubit system, it is simple to define operators that
project on to the spin-up or spin-down states. These projection
operators are: \be {\cal P}_0 \equiv \mid 0><0 \mid
\hspace{30pt}{\cal P}_1 \equiv \mid 1><1 \mid . \ee These are
Hermitian operators, and by virtue of closure, sum to one
$\nobreak{ \sum_{a=0,1} {\cal P}_a = 1.} $ They can also be
expressed in terms of the $\sigma_z$ operator as \be {\cal P}_0
=\frac{1 + \sigma_z}{2}  \qquad { \cal P}_1 =\frac{1 -
\sigma_z}{2}, \ee or in matrix form
 \be
  {\cal P}_0 = \left(
\begin{array}{lccc}
1 &&& 0 \\
0 &&& 0
\end{array}\right)
    \hspace{30pt}{ \cal P}_1 =
     \left(
\begin{array}{lccc}
 0 &&& 0 \\
0 &&& 1
\end{array}\right)\,.
\ee One can also project to other directions. For example,
projection of a qubit on to the $\pm\hat{x}$ or $\pm\hat{y}$
directions involves the projection operators \bea {\cal P}_{\pm x}
&=& \mid \pm \hat{x} > < \pm \hat{x} \mid
 = \frac{1 \pm \sigma_x}{2}, \nonumber \\
{\cal P}_{\pm y} &=& \mid \pm \hat{y} > < \pm \hat{y} \mid
= \frac{1 \pm \sigma_y}{2}.
\eea

\subsubsection{Usage}

The above projections operators are invoked in \QDENS by:

\bea
{\rm In[1] }&:=& {\rm \bf \mathcal{P}_0} \nonumber \\
{\rm Out[1]}&:=& {\bf \left(
\begin{array}{lccc}
 1 &&& 0 \\
 0 &&& 0
\end{array}
\right) }\nonumber
 \eea

\bea
{\rm In[2] }&:=& {\rm \bf \mathcal{P}_1} \nonumber \\
{\rm Out[2]}&:=& {\bf \left(
\begin{array}{lccc}
 0 &&& 0 \\
 0 &&& 1
\end{array}
\right) } \nonumber
 \eea Projection operators using the x-basis
are invoked by: \bea
{\rm In[3] }&:=& {\rm \bf {PX}[0]} \nonumber \\
{\rm Out[3]}&:=&  \left(
\begin{array}{lccc}
 \frac{1}{2} &&& \frac{1}{2} \\
 \frac{1}{2} &&& \frac{1}{2}
\end{array}
\right) \nonumber  \eea

\bea {\rm In[4] }&:=& {\rm \bf {PX}[1]}  \nonumber \\
 {\rm Out[4]}&:=&  \left(
\begin{array}{ll}
\  \ \frac{1}{2} & -\frac{1}{2} \\
 -\frac{1}{2} &\ \ \frac{1}{2}
\end{array}
\right)
 \nonumber
 \eea

A general operator {\bf ProjG[a,vec]} to project into a direction
stipulated by a unit three vector {\bf vec}, with a=0 or 1, is also
provided.  See {\it Tutorial.nb} for more examples. These operators
are useful for projective measurements.

\section{THE (SPIN) DENSITY MATRIX}
\label{sec3}

The above spin rotated wave functions can be used to obtain
the expectation value of some relevant Hermitian operator
$\Omega=\Omega^\dagger$, which represents a physical observable.
Let us assume that a system is in a state labelled by $\alpha$
with a state vector $\mid \alpha>$. In general the role of the
label $\alpha$ could be to denote a spatial coordinate
(or a momentum), if we were considering an ensemble of localized
particles. For the spin density matrix, we use $\alpha$ to
label the various spin directions $\hat{n}.$

The average or expectation value of a general observable $\Omega$
is then $ \nobreak{<\alpha \mid \Omega \mid \alpha >} \, . $ This
expectation value can be interpreted simply by invoking
eigenstates of the operator $\Omega$ \be \Omega \mid \nu > =
\omega_\nu \mid \nu >, \ee where $\omega_\nu$ are real eigenvalues
and $\mid \nu >$ are the eigenstates, which are usually a complete
orthonormal basis (CON). The physical meaning of the eigenvalue
equation is that if the system is in the eigenstate $\mid \nu >$,
there is no uncertainty $\Delta \Omega$ in determining the
eigenvalue, e.g. \be (\Delta \Omega)^2 \equiv <\nu \mid \Omega^2
\mid \nu> - < \nu \mid \Omega \mid \nu >^2 = \omega_\nu^2
-\omega_\nu^2 \equiv 0. \ee

Using the eigenstates $\mid \nu >$, we can now see the basic
meaning of the expectation value, which is a fundamental part of
QM. The eigenstates form a CON basis. That means any function can
be expanded in this basis and that the coefficients can be
obtained by an overlap integral. For example, in general terms the
completeness (C) allows the expansion \be \mid  \Psi> = \sum_\nu
c_\nu \mid \nu > . \ee The OrthoNormal (ON) aspect is $< \nu \mid
\nu'>=\delta_{\nu \nu'}.$ Thus \be <\nu' \mid \Psi>= \sum_\nu
c_\nu~<\nu'\mid~\nu>~=~c_{\nu'} \, , \ee reinserting this yields
\be \mid \Psi> = \sum_\nu <\nu \mid \Psi>  \mid \nu > = \sum_\nu \
\mid \nu ><\nu \mid \  \Psi> ={\bf I} \mid \Psi> . \nonumber \ee
Thus we see completeness with orthonormality of the basis can be
expressed in the closure form \be \sum_\nu \ \mid \nu><\nu\mid ={\bf
I} , \ee with ${\bf I}$ the unit operator in the Hilbert space.

With closure (e.g. a CON basis), we can now see that the
expectation value breaks in to a sum of the form
\bea
<\alpha \mid
\Omega \mid \alpha > &=& \sum_\nu
\sum_{\nu'} <\alpha \mid \nu>
<\nu\mid \Omega \mid \nu'>< \nu' \mid \alpha> \nonumber \\
&=& \sum_\nu \omega_\nu < \nu \mid \alpha><\alpha \mid \nu>
= \sum_\nu \omega_\nu \textit{P}_\nu^\alpha . \nonumber
\eea

Here
$\textit{P}^\alpha_\nu = < \nu \mid \alpha><\alpha \mid \nu>
= \mid< \nu \mid \alpha>\mid^2$ is the positive real probability
of the state $\mid \alpha>$ being in the eigenstate $\mid \nu>$.
Hence we see that the quantum average or expectation value is a
sum of that probability times the associated eigenvalue
$\omega_\nu$ over all possible values $\nu$. That is the
characteristic of a quantum average.

As the next step towards the spin density matrix, consider the
case that we have an ensemble of such quantum systems. Each system
is considered not to have quantum interference with the other
members of the ensemble. That situation can be realized by the
ensemble being located at separate sites with non-overlapping
localized wave packets and also in the case of a low density
beam, i.e. separated particles in the beam. This allows us to take
a classical average over the ensemble.

Suppose that the first member of the ensemble is produced in
the state $\mid \alpha >$, the next in $\mid \alpha' >$, etc.
The ensemble average is then a simple classical average
\be
<\Omega> =  \frac{\sum_{\alpha} <\alpha \mid \Omega \mid \alpha >
{\bf P}_\alpha}{\sum_{\alpha} {\bf P}_\alpha},
\ee
where ${\bf P}_\alpha$ is the probability that a particular
state $\alpha$ appears in the ensemble. Summing over all
possible states of course yields $\sum_{\alpha} {\bf P}_\alpha=1$.
The above expression is a combination of a classical ensemble
average with the quantum mechanical expectation value. It
contains the idea that each member of the ensemble interferes
only with itself quantum mechanically and that the ensemble
involves a simple classical average over the probability
distribution of the ensemble.

We are close to introducing the density matrix. This is
implemented by using closure and rearranging. Consider
\be
\sum_{\alpha} <\alpha \mid \Omega \mid \alpha > {\bf P}_\alpha
=
\sum_{\alpha}\ \sum_{m m'} <\alpha \mid m >
 < m \mid \Omega \mid m'><m' \mid \alpha>\ {\bf P}_\alpha \,,
\ee
where $\mid m >$ denotes any CON basis. Now rearrange the
above to
\be
\sum_{\alpha} <\alpha \mid \Omega \mid \alpha > {\bf P}_\alpha
= \sum_{m m'} \sum_{\alpha}\ <m'\mid \alpha><\alpha \mid m> \,
{\bf P}_\alpha < m \mid \Omega \mid m'> \,,
\ee
and then define the {\bf density operator} by
\be
\rho \equiv \sum_{\alpha}\ \mid \alpha><\alpha \mid {\bf P}_\alpha
\ee
and the associated density matrix in the CON basis
$\mid m>$
as
$<m \mid \rho \mid m'> = \sum_{\alpha}\ < m \mid \alpha>
<\alpha \mid m'> {\bf P}_\alpha$.
We often refer to either the density operator or the density
matrix simply as the ``density matrix," albeit one acts in the
Hilbert space and the other is an explicit matrix. The ensemble
average can now be expressed as a ratio of
traces~\footnote{The trace ${\rm Tr}$ is defined as the sum
of the diagonal matrix elements of an operator, where a CON
basis is used.}
\be
<\Omega>= \frac{{\rm Tr}[ \rho \Omega]}{{\rm Tr}[ \rho ]},
\ee
which entails the properties that
\bea
{\rm Tr} [\rho] &=&  \sum_m <m\mid \rho\mid m>
= \sum_\alpha\ {\bf P}_\alpha \sum_m  <\alpha \mid m><m\mid \alpha>\nonumber \\
&=& \sum_\alpha\ {\bf P}_\alpha   <\alpha \mid  \alpha>
= \sum_\alpha\ {\bf P}_\alpha=1,
\eea
and
\bea
{\rm Tr} [\rho \Omega]  &=&
\sum_{m m'}  <m\mid \rho \mid m'><m'\mid \Omega\mid m> \nonumber \\
&=& \sum_\alpha \sum_{m m'} {\bf P}_\alpha <\alpha\mid m'>
<m' \mid \Omega\mid m><m\mid \alpha> \nonumber \\
&=&\sum_\alpha {\bf P}_\alpha <\alpha\mid \Omega \mid \alpha> ,
\eea
which returns the original ensemble average expression.

\subsection{Properties of the Density Matrix}

We have  defined the density operator by a sum involving state
labels $\alpha$ for the special case of a spin 1/2 system. The
definition \be \rho = \sum_{\alpha}\ \mid \alpha><\alpha \mid {\bf
P}_\alpha \ee is however a general one, if we interpret $\alpha$
as the label for the possible characteristics of a state. Several
important general properties of a density operator can now be
delineated. The density matrix is Hermitian, hence its eigenvalues
are real. The density matrix is also positive definite, which
means that all of its eigenvalues are greater or equal to zero.
This, together with the fact that the density matrix has unit
trace, ensures that the eigenvalues are in the range [0,1].

To prove that the density matrix is positive definite, consider
a basis $\mid \nu>$ which diagonalizes the density operator
so that
\bea
< \nu \mid \rho \mid \nu> &=&  \mu_\nu  \\
&=& \sum_\alpha  {\bf P}_\alpha  <\nu\mid \alpha><\alpha \mid \nu>
= \sum_\alpha  {\bf P}_\alpha  \mid < \nu \mid \alpha>\mid^2  \geq
0. \nonumber \eea
Here $\mu_\nu$ is the $\nu$th eigenvalue of
$\rho$ and both parts of the final sum above are positive
quantities. Hence all of the eigenvalues of the density matrix are
$\geq 0$ and the density matrix is thus positive definite. If one
of the eigenvalues is one, all the others are zero.

Another general property of the density matrix involves the
special case of a pure state. If every member of the ensemble has
the same quantum state, then only one $\alpha$ (call it $\alpha_0$)
appears and the density operator becomes $\nobreak{\rho
= \mid \alpha_0><\alpha _0 \mid}$. The state $\mid \alpha_0> $ is
normalized to one and hence for a pure state $\rho^2 = \rho$.
Using a basis that diagonalizes $\rho$, this result tells us that
the eigenvalues satisfy $\mu_\nu ( \mu_\nu-1) =0$ and hence for a
pure state one density matrix eigenvalues is 1, with all others
zero.

In general, an ensemble does not have all of its members in
the same state, but has a mixture of possibilities as reflected
in the probability distribution ${\bf P}_\alpha$. In general,
as we show below, we have
\be
\rho ^2 \leq \rho,
\ee
with the equal sign holding for pure states. A simple way to
understand this relationship is seen by transforming the
density matrix to diagonal form, using its eigenstates to
form a unitary matrix $U_\rho$. We have
$U_\rho \rho U^\dagger_\rho = \rho_D,$ where $\rho_D$ is diagonal
using the eigenstates of $\rho$ as the basis, e.g.
$<\nu \mid \rho_D \mid \nu'>= \mu_\nu \delta_{\nu \nu'}$.
Here $\mu_\nu$ again denotes the $\nu$th eigenvalue of $\rho$. We
already know that the sum of all these eigenvalue equals 1, that
they are real and positive. Since every eigenvalue is limited
to be less than or equal to 1, we have $\mu_\nu^2\leq \mu_\nu$,
for all $\nu$. Transforming that back to the original density
matrix yields the result $\rho^2 \leq \rho$. Taking the trace
of this result yields another test for the purity of the state
${\rm Tr}[\rho^2] \leq {\rm Tr}[\rho]=1$. Examples of how to use
this measure of purity will be discussed later.

\subsubsection{Entropy and Fidelity}

As an indication of the rich variety of functionals of $\rho$ that
 can be defined,  let us examine the Von
Neumann entropy and the fidelity.

The Von Neumann entropy~\cite{Neumann}, $S[\rho]=-{\rm Tr} [\rho\;
\log_2 \rho]$, is a measure of the degree of disorder in the
ensemble. Its basic properties are: $S[\rho]=0$ if $\rho$ is a
pure state, and $S[\rho]=1$ for completely disordered states. See
later for an application to the Bell,  GHZ , \& Werner states and
also the {\it Tutorial.nb} notebook for simple illustrative
examples.

It is often of interest to compare two different density matrices
that are alternate descriptions of an ensemble of quantum systems.
One simple measure of such differences is the fidelity. Consider
for example,  two pure states \be \rho = \mid \psi > <\psi \mid
\hspace{.25in}\rho = \mid \widetilde{ \psi }
> < \widetilde{ \psi  } \mid  ,
\ee
and the associated overlap obtained by a trace method
\be {\rm Tr}[\ \rho\ \widetilde{\rho} \ ] =
{\rm Tr}[ \ \mid \psi
> < \psi \mid
 \ \widetilde{ \psi} > < \widetilde{ \psi} \mid  \ ] =
 |< \psi \mid \widetilde{ \psi}> |^2
 .
\ee Clearly this overlap equals one if the states and associated
density matrices are the same and thus serves as a measure to
compare states. This procedure is generalized and applied to
general density matrices. It is also written in a symmetric
manner, with the general definition of fidelity being
 \be
 F[ \rho, \widetilde{\rho} ] = {\rm Tr}[ \sqrt{ \sqrt{\tilde{\rho}}
 \ \rho \  \sqrt{\tilde{\rho} }  \
  } ] \ ,
 \ee
which has the property of reducing to ${\rm Tr}[ \rho ]=1,$ for $\tilde{\rho}=\rho.$ It
also yields $ |< \psi \mid \widetilde{ \psi}> | $ in the pure state limit.

\subsubsection{Usage}

 \QDENS includes commands that produce the Purity and Entropy for a
 stipulated density matrix $\rho,$ {\bf
\Purity[$\rho$], \Entropy[$\rho$],} and the Fidelity of one
specified density matrix  relative to another
\Fidelity[$\rho_1,\rho_2$].

\subsubsection{ Composite Systems and Partial Trace}

For a composite system, such as colliding beams, or an ensemble of quantum systems
each of which is prepared with a probability distribution,
 the definition of a density matrix can be generalized to a
 product Hilbert space form involving systems of type A or B
\be \rho_{A B} \equiv  \sum_{\alpha, \beta}  {\bf P}_{\alpha,
\beta} \mid \alpha \beta>< \alpha \beta \mid, \ee  where ${\bf
P}_{\alpha, \beta}$  is the joint probability for finding the two
systems with the attributes labelled by $ \alpha$ and $\beta.$ For
example, $\alpha$ could designate the possible directions
$\hat{n}$ of one spin-1/2 system, while $\beta$ labels the
possible spin directions of another spin 1/2 system.  One can
always ask about the state of system A or B by summing over or
tracing out the other system. For example the density matrix of
system A is picked out of the general definition above by the following trace steps
 \bea
 \rho_A &=&  {\rm Tr}_B [  \rho_{A B} ]  \nonumber \\
  &=&  \sum_{ \alpha, \beta} \  {\bf P}_{ \alpha, \beta} \mid \alpha> < \alpha\mid
  {\rm Tr}_B[ \ \mid \beta>< \beta \mid \ ]  \nonumber \\
&=&  \sum_{ \alpha}  ( \sum_{ \beta}  {\bf P}_{\alpha, \beta} )
\mid \alpha> < \alpha\mid
   \nonumber \\
   &=&  \sum_{\alpha}  {\bf P}_{\alpha}  \mid \alpha> < \alpha\mid .
 \eea
Here we use the product space $ \mid \alpha \beta>\mapsto \mid \alpha> \mid \beta>$ and
we define the probability for finding system A in situation $\alpha$ by
\be
{\bf P}_{\alpha} =\sum_{ \beta}  {\bf P}_{\alpha, \beta}.
\ee
This is a standard way to get an individual probability from a joint probability.

It is easy to show that all of the other properties of a density matrix still hold
true for a composite system case. It has unit trace, it is Hermitian with real
eigenvalues, etc.

See later for application of these general properties to multi-qubit systems.

\subsection{Comments about the Density Matrix}

\subsubsection{Alternate Views of the Density Matrix}

In the prior discussion, the view was taken that the density
matrix implements a classical average over an ensemble of many
quantum systems, each member of which interferes quantum
mechanically only with itself. Another viewpoint, which is equally
valid, is that a single quantum system is prepared, but the
preparation of this single system is not pinned down. Instead all
we know is that it is prepared in any one of the states labelled
again by a generic state label $\alpha$ with a probability
${\bf P}_{\alpha}$. Despite the change in interpretation, or
rather in application to a different situation, all of the
properties and expressions presented for the ensemble average
hold true; only the meaning of the probability is altered.

Another important point concerning the density matrix is that
the ensemble average (or the average expected result for a
single system prepared as described in the previous paragraph)
can be used to obtain these averages for all observables $\Omega$.
Hence in a sense the density matrix describes a system and
the system's accessible observable quantities. It represents then
an honest statement of what we can really know about a system. On
the other hand, in Quantum Mechanics it is the wave function
that tells all about a system. Clearly, since a density matrix
is constructed as a weighted average over bilinear products
of wave functions, the density matrix has less detailed information
about a system that is contained in its wave function. Explicit
examples of these general remarks will be given later.

To some authors the fact that the density matrix has less
content than the system's wave function, causes them to avoid
use of the density matrix. Others find the density matrix description
of accessible information as appealing.

\subsubsection{ Classical Correlations and Entanglement}

The density matrix for composite systems can take many forms depending
 on how the systems are prepared.  For example, if distinct systems A \& B are
  independently produced and observed independently,  then the density
  matrix is of product form
$\rho_{AB} \mapsto \rho_A \otimes \rho_B,$ and the observables are also
 of product form
$\Omega_{AB} \mapsto\Omega_A\otimes\Omega_B.$  For such an uncorrelated
situation,  the ensemble
average factors
\be
< \Omega_{AB}> = \frac{{\rm Tr}[ \rho_{AB}\Omega_{AB}]}
{{\rm Tr}[ \rho_{AB}]}= \frac{{\rm Tr}[ \rho_{A}\Omega_{A}]}{{\rm Tr}[ \rho_{A}]}
 \frac{{\rm Tr}[ \rho_{B}\Omega_{B}]}{{\rm Tr}[ \rho_{B}]}
\ee  as is expected for two separate uncorrelated experiments.
This can also be expressed as having the joint probability factor
${\bf P}_{\alpha, \beta}\mapsto{\bf P}_{\alpha  }  {\bf P}_{ \beta}$
the usual probability rule for uncorrelated systems.

Another possibility for the two systems is that they are prepared in a
coordinated manner,
with each possible situation assigned a probability based on the
correlated preparation technique.  For example, consider two colliding beams, A \& B,
made up of particles with the same spin. Assume the particles are produced
in matched pairs with common spin direction $\hat{n}.$ Also
assume that the preparation of that pair in that shared direction
is produced by design with a
 classical probability distribution
${\bf P}_{\hat{n}}.$  Each pair has a density matrix $
\rho_{\hat{n}} \otimes \rho_{\hat{n}}$ since they are produced
separately,  but their spin directions are correlated classically.
The density matrix for this situation is then
\be \rho_{AB} =
\sum_{\hat{n}} {\bf P}_{\hat{n }} \  \rho_{\hat{n}} \otimes
\rho_{\hat{n}}.
\ee
This is a ``mixed state" which represents
classically correlated preparation and hence any density matrix
that takes on the above form can be reproduced by a setup using
classically correlated preparations and does {\it not} represent
the essence of Quantum Mechanics an entangled state.

An entangle quantum state is described by a density matrix (or by
its corresponding state vectors) that is not and can not be
transformed into the two classical forms above; namely, cast into
a product or a mixed form. For example, a Bell state
$\frac{1}{2} (\mid 01 > + \mid 10>)$ has a density matrix
\be
\rho= \frac{1}{2}( \mid 01> < 01 \mid + \mid 01> < 10 \mid
                + \mid 10> < 01 \mid + \mid 10> < 10 \mid \, )
\ee
that is not of simple product or mixed form. It is the prime
example of an entangled state.

The basic idea of decoherence can be described by considering the
above case with time dependent coefficients
\be
\rho= \frac{1}{2}(a_1(t) \mid 01> < 01 \mid +a_2(t) \mid 01> < 10 \mid +a_2^*(t)
\mid 10> < 01 \mid +a_3(t) \mid 10> < 10 \mid \, ) .
\ee If the
off-diagonal terms $a_2(t)$ vanish, by attenuation and/or via
time averaging, then the above density matrix does reduce to the
mixed or classical form, which is an illustration of how
decoherence leads to a classical state.

\section{MULTI -QUBIT SYSTEMS}
\label{sec4}

The previous discussion which focused on describing a single
qubit, can now be generalized to multiple qubits. Consider the
product space of two qubits both in the up state and denote that
product state as $ \mid 0\ 0>= \mid 0> \mid 0>, $ which clearly
generalizes to
\be
\mid q_1\ q_2>= \mid q_1> \mid q_2>,
\ee
where $q_1,q_2$ take on the values $0$ and $1$. This product is called a
tensor product and is symbolized as
\be
\mid q_1\ q_2>= \mid
q_1>\otimes \mid q_2>, \ee which generalizes to $n_q$ qubits \be
\mid q_1\ q_2\ \cdots\ q_{n_q}>= ( \mid q_1>\otimes \mid
q_2>)\, ( \cdots \otimes \mid q_{n_q}>) .
\ee

\begin{figure}[t]
\fbox{\parbox{1.\textwidth}{
\includegraphics[width=8pc]{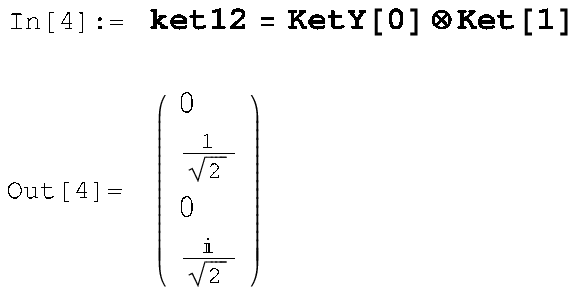}
\hspace*{0.5cm}
\includegraphics[width=12pc]{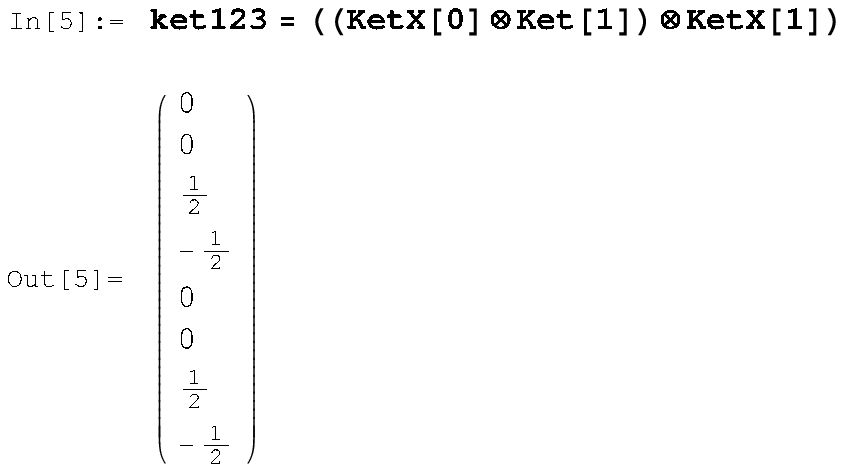}
\hspace*{0.5cm}
\includegraphics[width=8pc]{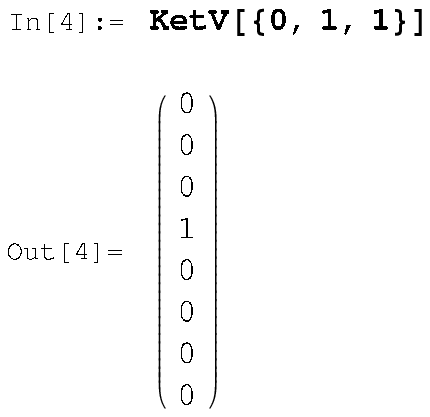}
}}
\caption{Simple examples of tensor products of two and three kets.}
\protect\label{kets}
\end{figure}

In \QDENS, the kets $\mid 0>,\mid 1>$ are invoked by the commands
\Ket[0] and \Ket[1], as shown in Fig.~\ref{kets}, along with the
kets $\mid \pm x>$, and $ \mid \pm y>$.

Also shown in that figure are the results for forming the tensor
products of the kets for two and three qubits, as described next.

\subsection{Multi-Qubit Operators}

\begin{figure}[t]
\fbox{\parbox{0.7\textwidth}{\includegraphics[width=10cm]{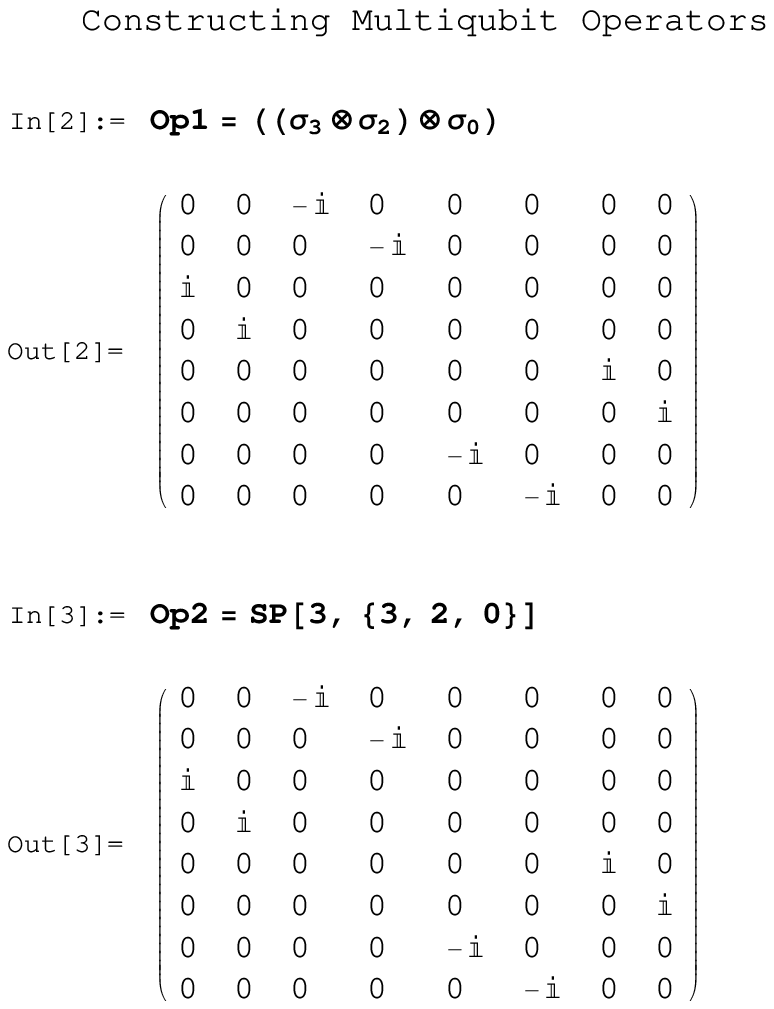}}}
 \caption{Multi-qubit operators in \QDENS}
\protect\label{multiqubops}
\end{figure}

One can also build operators that act in the multi-qubit spin
space described above. Instead of a single operator, we have a set
of separate Pauli operators acting in each qubit space. They
commute because they refer to separate, distinct quantum systems.
Hence, we can form the tensor product of the $n_q$ spin operators
which for two qubits has the following structure \bea < a_1 \mid
\sigma_i \mid b_1> < a_2 \mid \sigma_j \mid b_2>&=&
< a_1 a_2  \mid \sigma^{(1)}_i \sigma^{(2)}_j \mid b_1 b_2> \nonumber \\
&=&
< a_1 a_2  \mid \sigma^{(1)}_i  \otimes  \sigma^{(2)}_j \mid b_1 b_2>\,,
\eea
which defines what we mean by the tensor product $\sigma^{(1)}_i \otimes \sigma^{(2)}_j$
for two qubits. The generalization is immediate
\be
( \sigma^{(1)}_i  \otimes  \sigma^{(2)}_j )  \otimes( \sigma^{(3)}_k
\otimes  \sigma^{(4)}_l) \cdots \ .
\ee
The corresponding steps in \QDENS are shown in Fig.~\ref{multiqubops},

For large numbers of qubits a recursive method has been developed
(see, {\it Qdensity.m} and {\it Tutorial.nb}), which involves
specifying the ``Length" L$= n_q$ of the qubit array and an array
of length L that specifies the Pauli components ${i,j,k,l,
\cdots}$. For example, if $i=1, j=0$ there is a $\sigma_x$ in
qubit 1 space and a unit operator $\sigma_0$ acting in qubit 2
space. The multi-qubit spin operator is called ${\rm SP}[L,
\{i,j,k,l, \cdots\} ]$. Examples  in Fig.~\ref{multiqubops}
include operator tensor products generated  directly using the
$\otimes$ notation.

\subsubsection{Usage}

\QDENS includes a  multiqubit spin operator {\bf
SP[L,\{$a_1$,$a_2$,..,$a_L$\}] } built from $L$ Pauli spin operators
of components $a_1$,$a_2$,..,$a_L.$  A sample construction is:

\bea {\rm In[1] }&:=& {\rm \bf {SP}[2,\{2,3\}]}  \nonumber \\
 {\rm Out[1]}&:=&  \left(
\begin{array}{llll}
\ \ 0 &\ \  0 & -i &\ \ 0 \\
 \ \ 0 & \ \ 0 & \ \ 0 &\  \ i \\
\ \ i & \ \ 0 &\ \  0 & \ \ 0 \\
\  \ 0 & -i & \ \ 0 & \ \ 0
\end{array}
\right)
 \nonumber
 \eea which is equivalent to the tensor product

 \bea {\rm In[2] }&:=& {\rm \bf \sigma _2 \otimes  \sigma _3}  \nonumber \\
 {\rm Out[2]}&:=&  \left(
\begin{array}{llll}
 \ \ 0 &\ \  0 & -i &\ \ 0 \\
 \ \ 0 & \ \ 0 & \ \ 0 &\  \ i \\
\ \ i & \ \ 0 &\ \  0 & \ \ 0 \\
\  \ 0 & -i & \ \ 0 & \ \ 0
\end{array}
\right)
 \nonumber
 \eea
 The advantage of this command is that it can readily construct large space tensor
 products.

\subsection{General Multi -Qubit Operators}

The production of $n_q$ spin space operators provides a complete
basis for expressing any operator. This remark is similar to, and
indeed equivalent to, the statement that the $n_q$ ket product
space is a CON basis. With that remark, we can expand any $n_q$
operator as \be \Omega = \sum_{\bf a} {\cal C}_{ \bf a} \;
\sigma^{(1)}_{a_1} \otimes \sigma^{(2)}_{a_2} \otimes
\sigma^{(3)}_{a_3}\cdots \sigma^{(n_q)}_{a_{n_q}} = \sum_{\bf a}
{\cal C}_{ \bf a} \ {\rm SP}[n_q,{\bf a}] \ee where the sum is
over all possible values of the array ${\bf a}:\{
a_1,a_2,a_3,\cdots, a_{n_q} \}$. Here, the multi-qubit spin
operator is denoted by \SP$[n_q,{\bf a}]$, which is the notation
used in \QDENS. The coefficient ${\cal C}_{\bf a}$ can be evaluated
for any given $\Omega$ from the overall trace \be {\cal C}_{\bf a}
= \frac{1}{2^{n_q}}  {\rm Tr}[\, \Omega\ .{\rm SP}[n_q,{\bf a}]\,  ].
\ee Because of the efficacy of Mathematica 5.2, the total trace
can be evaluated rapidly. This set of coefficients characterizes
the operator $\Omega$.

\subsubsection{Partial Traces}

The advantage of expanding a general operator in the Pauli
operator basis is that partial traces can now be generated by
manipulating the above coefficients. A partial trace involves
tracing out parts of a system; for example, consider the partial
trace over qubit two for a three qubit operator \be {\rm Tr}_2 [
\sigma^{(1)}_i \otimes \sigma^{(2)}_j \otimes \sigma^{(3)}_k]= 2
\delta_{j 0} \  \sigma^{(1)}_i \otimes \sigma^{(3)}_k\ . \ee
Recall that ${\rm Tr}[\sigma_i]$ is zero unless we have the unit
matrix $\sigma_0$ in which case the trace is two. Of course, one
could trace out systems 2 and also 1, and then \be {\rm Tr}_{1 2}
[ \sigma^{(1)}_i \otimes \sigma^{(2)}_j \otimes \sigma^{(3)}_k]= 2
\delta_{i 0}\ 2 \delta_{j 0}\ \sigma^{(3)}_k\ . \ee The subscript
on the {\rm Tr} symbol indicates which qubit operators are being
traced out. Note in this case the number of qubits in the result
is reduced to $n_q-2$, where 2 is the length of the subscript
array in ${\rm Tr}_{1 2}$. Clearly, the trace reduces the
rank~\footnote{The rank is the number of qubits $n_q$.} of the
operator by the number of qubits traced out.

Now we can apply these simple ideas to construct the partial trace
of a general operator $\Omega$. Using the linearity of the trace
\be {\rm Tr}_{\bf t} [ \Omega ] =\sum_{\bf a} {\cal C}_{ \bf a}\
{\rm Tr}_{\bf  t}[\sigma^{(1)}_{a_1} \otimes \sigma^{(2)}_{a_2}
\otimes \sigma^{(3)}_{a_3}\cdots \ \sigma^{(n_q)}_{a_{n_q}}], \ee
where the array ${\bf t}:\{t_1 t_2 \cdots \}$ indicates only those
qubits that are to be traced out. For example, ${\bf t}:\{2 5 \}$
indicates that only qubits 2 and 5 are traced out.

\begin{figure}[t]
\fbox{\parbox{0.8\textwidth}{\includegraphics[width=10cm]{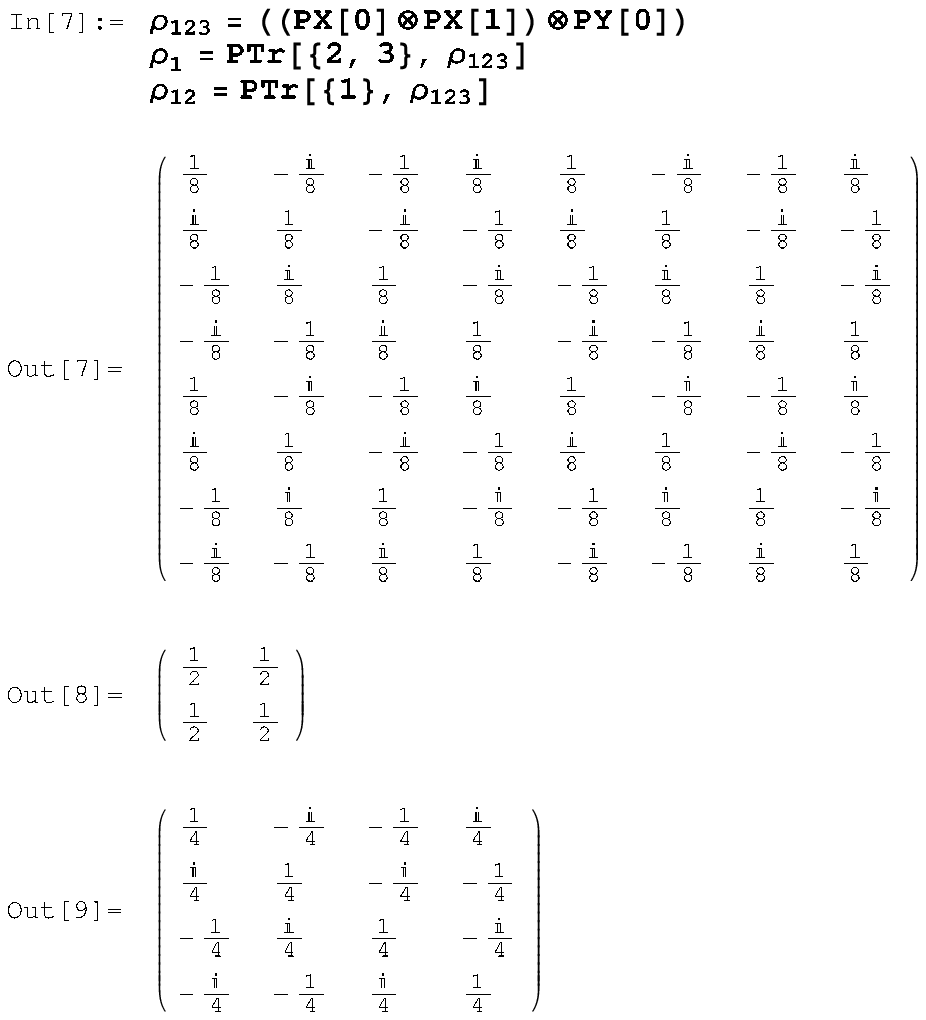}}}
 \caption{Taking partial traces with \QDENS}
\protect\label{traceout1}
\end{figure}

The procedure for taking a partial trace of a general operator is
to determine the total coefficient ${\cal C}_{\bf a}$ for all of
the array ${\bf a}:\{a_1 a_2 \cdots a_{n_q} \},$ except for the
entries corresponding to the traced out qubit for which we need
only the $a_j=0$ part if say we trace out the $j$th qubit. From
the resultant coefficients, we obtained a reduced set of
coefficients, reduced by the number of trace outs. That reduced
coefficient is then used to construct the reduced space operator,
with a multiplier of 2 included for each traced out qubit. This
expansion, reduction, reconstruction procedure might seem
complicated, but it has been implemented very efficiently using
the power of Mathematica 5.2. See {\it Qdensity.m} for the explicit
construction procedure (which is rather compact). The command
used in \QDENS is \PTr[ {\bf t}, $\Omega$] where the trace out of
the general operator $\Omega$ is specified by the array
$ {\bf t}$. Examples of the partial traces are  in
Fig.~\ref{traceout1}.

\subsubsection{Usage}
 \QDENS includes several commands for taking partial traces. One is {\bf
 PTr[\{$q_1$,$q_2$,...,$q_M$\},$\Omega$]},  where the array
 $q_1$,$q_2$,...,$q_M$ stipulates the
 space to be traced out.  See {\it Tutorial.nb} and Fig.~\ref{traceout1} for  examples
 of these commands.

\subsection{Multi-Qubit Density Matrix }

\begin{figure}[t]
\fbox{\parbox{0.7\textwidth}{\includegraphics[width=8cm]{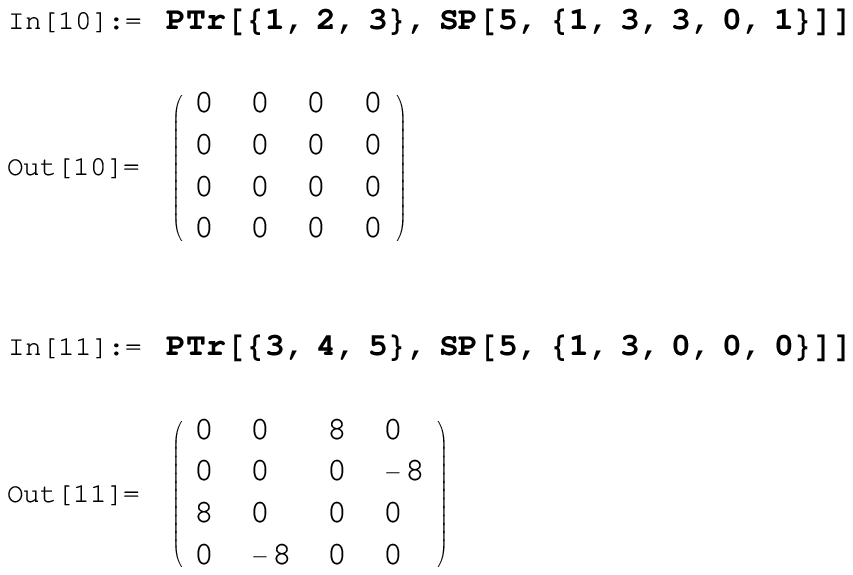}}}
\caption{Partial Traces of multi-qubit Pauli operators}
\protect\label{ptracerho}
\end{figure}

\begin{figure}[t]
\fbox{\parbox{.8\textwidth}{
\includegraphics[width=10cm]{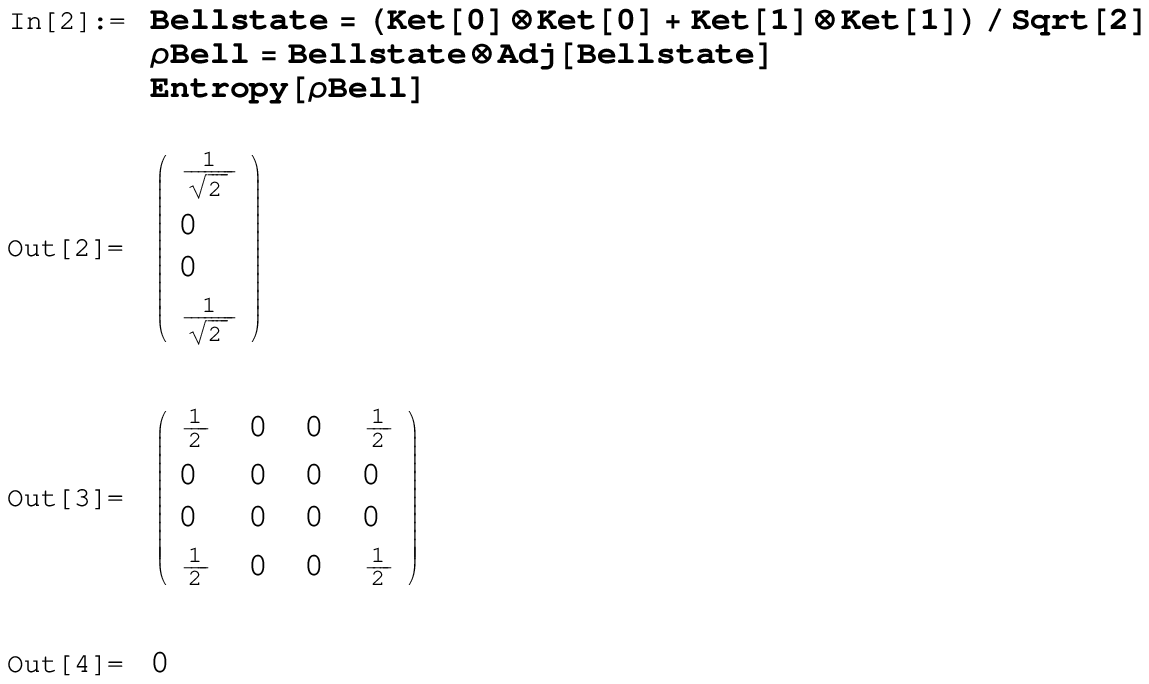}
\includegraphics[width=10cm]{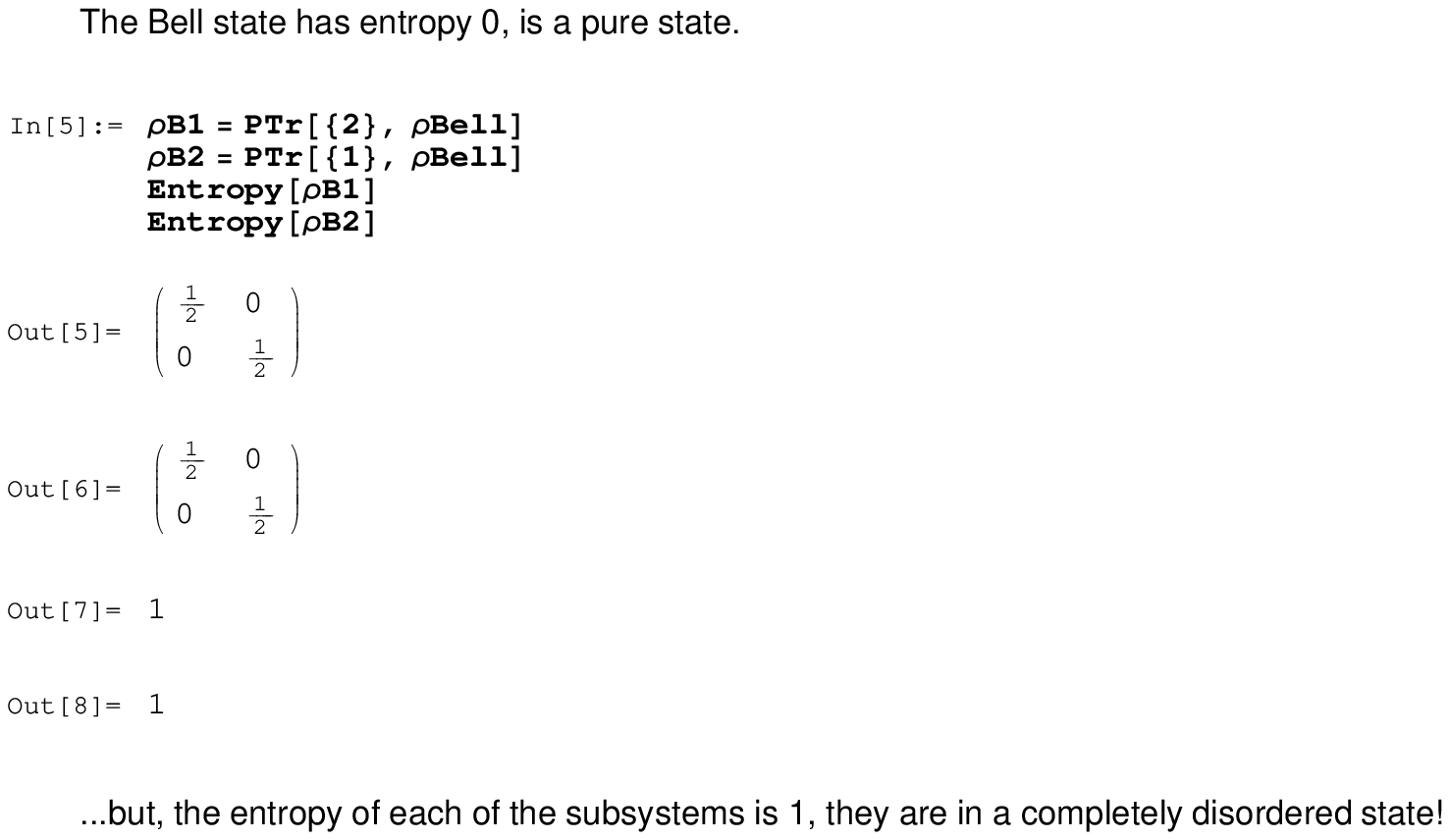}
}}
 \caption{Example from {\it Tutorial.nb}.}
\protect\label{Bellreduce}
\end{figure}

The multi-qubit density matrix is our prime example of an operator
that we examine in various ways, including taking partial traces.
Just as in the prior discussion, a general density matrix can be
expanded in a Pauli spin operator basis \be \rho = \sum_{\bf a} \
{\cal C_\rho}_{ \bf a}\ \ {\rm SP}[n_q,{\bf  a}], \ee where the
coefficient ${\cal C_\rho}_{ \bf a}$ is real since the density
matrix and the Pauli spin tensor product \SP$[n_q,{\bf  a}]$ are
Hermitian. Taking a partial trace follows the rules discussed
earlier. Examples are presented in Figs.~\ref{ptracerho} and~\ref{Bellreduce}.

In these examples, we give the case of three qubits reduced to two
and then to one. The general expansion for these cases takes on a
simple and physically meaningful form and therefore are worth
examining. For one qubit, the above expansion is of the
traditional form \be \rho_1 = \frac{1}{2} [ \ {\bf 1} +
\vec{P}_1\cdot \vec{\sigma}], \ee which involves the three numbers
contained in the vector $\vec{P}_1$, also know as the polarization
of the ensemble. A $2\times2$ Hermitian density matrix has 4
variables, which is reduced by one by the ${\rm Tr}[\rho_1]=1$
normalization. Thus the polarization vector is a complete
parametrization of a single qubit. For a pure state, the magnitude
of the polarization vector is one; whereas, the general constraint
$\rho^2 \leq \rho$ implies that $\mid P_1 \mid \leq 1$. A graph of
that vector thus lies within a unit circle called the Bloch
sphere. The physical meaning of $\vec{P}_1$ is that it is the
average polarization of an ensemble, which is made clear by
forming the ensemble average of the Pauli spin vector: \be <
\vec{\sigma}> = \frac{{\rm Tr} [ \rho_1 \vec{\sigma} ]}{ {\rm
Tr}[\rho_1]} \equiv \vec{P}_1. \ee

Now consider two qubits. The Pauli basis is $\sigma_i \otimes \sigma_j$, and
hence the two qubit density matrix has the form
\bea
\rho_{1 2} &=& \frac{1}{4} [ \ {\bf 1} + \vec{P}_1 \cdot \vec{\sigma}_1 \otimes {\bf 1}
+ {\bf 1} \otimes \vec{\sigma}_2 \cdot \vec{P}_2 +  \sigma_{1 i} \otimes
\sigma_{2 j} T_{i , j} ] \nonumber \\
&=& \frac{1}{4} [ \ {\bf 1} + \vec{P}_1 \cdot
\vec{\sigma}_1
 +
\vec{P}_2 \cdot \vec{\sigma}_2 + \vec{\sigma}_{1} \cdot \overleftrightarrow{ T}
\cdot \vec{\sigma}_{2 j}  ]. \nonumber
\eea
This involves two polarization vectors, plus one $3\times 3$ tensor polarization
$\overleftrightarrow{ T}$~\footnote{In the tensor term the sum extends only over
the $i,j=1,2,3$ components} which comes to 15 parameters as indeed is the correct
number for a two qubit system $ 2^2 \times 2^2 -1$~\footnote{We see that the number
of parameters in a $n_q$ qubit density matrix is thus
$ 2^{n_q} \times 2^{n_q}-1=2^{ 2 n_q} -1 .$}. The physical meaning is again an ensemble
average polarization vector for each qubit system, plus an ensemble average spin
correlation tensor
\bea
< \vec{\sigma}_1 > &=& \frac{{\rm Tr} [ \rho_{1 2}\ \vec{\sigma}_1
\otimes {\bf 1}_2 ]}{ {\rm Tr}[\rho_{1 2}]}\
\equiv \vec{P}_1, \nonumber \\
< \vec{\sigma}_2 > &=& \frac{{\rm Tr} [ \rho_{1 2}\
{\bf 1}_1 \otimes \vec{\sigma}_2 ]}{ {\rm Tr}[\rho_{1 2}]}
\equiv \vec{P}_2, \nonumber \\
< \sigma_{1 i} \sigma_{2 j} > &=& \frac{{\rm Tr} [ \rho_{1 2}\
\sigma_{1 i} \otimes \sigma_{2 j} ]}{ {\rm Tr}[\rho_{1 2}]}
\equiv T_{i j}.\nonumber \\
&&
\eea

To illustrate a partial trace, consider the trace over qubit 2 of
the two qubit density matrix
\be
{\rm Tr}_2[ \rho_{ 1 2}] = \rho_{ 1}=\frac{1}{2} [ {\bf 1} +
\vec{P}_1 \cdot \vec{\sigma}_1 ],
\ee
where we see that a proper reduction to the single qubit space
results. Examples of the density matrix for the Bell states and
their partial trace reduction to the single qubit operator are
presented in Fig.~\ref{Bellreduce}.

\subsection{Multi-Qubit States }

The procedure for building multi-qubit states follows a path
similar to our discussion of operators. First we build the
computational basis states, which are eigenstates of the operator
$\sigma^{(1)}_z \otimes \sigma^{(2)}_z\otimes \cdots
\sigma^{(n_q)}_z$. These states are specified by an array ${\bf
a}:\{ a_1, a_2 \cdots a_{n_q}\}$ of length $n_q$, where the
entries are either one or zero. That collection of binary bits
corresponds to a decimal number according to the usual rule $
a_1a_2 \cdots a_{n_q} \rightarrow a_1 2^{n_q} + a_2  2^{n_q-1} +
a_{n_q}  2^{0}$. The corresponding product state $ \mid a_1  a_2
\cdots a_{n_q}> \equiv\ \mid a_1 > \otimes \mid a_2> \otimes
\cdots \mid a_{n_q}> $ can be constructed using the command
{\goodfontB KetV[\{$a_1$, $a_2$, ..\}]}. Any single computational basis state
consists of a column vector with all zeros except at the location
counting down from the top corresponding to its decimal
equivalent. Examples of the construction of multiqubit states in
\QDENS are given in Fig.~\ref{kets}. This capability allows one to
use \QDENS without invoking a density matrix approach, which is
often desirable to reduce the space requirements imposed by a full
density matrix description.


\subsubsection{Usage}

\QDENS includes commands for one qubit ket vectors in the
computational and in the x- and y-basis  ${\bf Ket, KetX, KetY},$
and also multiqubit product states using the command {\bf
KetV[vec]}.   Example of its use is

\bea
{\rm In[1] }&:=& {\rm \bf \text{KetV}[\{0,1,1\}] } \nonumber \\
{\rm Out[1]}&:=& {\bf \left(
\begin{array}{l}
 0 \\
 0 \\
 0 \\
 1 \\
 0 \\
 0 \\
 0 \\
 0
\end{array}
\right) }\nonumber
 \eea
which is equivalent to

\bea {\rm In[1] }&:=& {\rm \bf  (\text{Ket}[0]\otimes
\text{Ket}[1])\otimes
\text{Ket}[1] } \nonumber \\
{\rm Out[1]}&:=& {\bf \left(
\begin{array}{l}
 0 \\
 0 \\
 0 \\
 1 \\
 0 \\
 0 \\
 0 \\
 0
\end{array}
\right) }\nonumber
 \eea

\section{CIRCUITS \& GATES}
\label{sec5}

Now that we can construct multi-qubit operators and take the partial trace,
we are ready to examine the operators that correspond to logical gates for
single and multi-qubit circuits. These gates form the basic operations that
are part of the circuit model of QC. We will start with one qubit operators
in a one qubit circuit and then go on to one qubit operators acting on selected
qubits within a multi-qubit circuit. Then two qubit operators in two and multi-qubit
situations will be presented.

\subsection{One Qubit Gates}
\subsubsection{NOT}
The basic operation \NOT is simply represented by the $\sigma_x$
matrix since $\nobreak{\sigma_x \mid 0> = \mid 1 >}$ and $
\sigma_x \mid 1
> = \mid 0>.$

\subsubsection{The Hadamard}
For the Hadamard, we have the simple relation
\be
{\cal H} = \frac{ \sigma_x + \sigma_z}{\sqrt{2}} \rightarrow
 \frac{1}{\sqrt{2} } \left(
\begin{array}{lcr}
1 && 1\\
1 && -1
\end{array}\right) ,
\ee
which can also be understood as a rotation about the
$\hat{\eta} =\frac{ \hat{x} + \hat{z}}{\sqrt{2}}$
axis by $\gamma=\pi$ since
\be R= e^{-\frac{\gamma}{2} \vec{\sigma}\cdot
\hat{\eta} }
 = \cos \frac{ \pi}{2}\;\sigma_0 - i \sin\frac{ \pi}{2}\; \vec{\sigma}\cdot \hat{\eta}
   \rightarrow -i \frac{ \sigma_x + \sigma_z}{\sqrt{2}}
   \label{rot} \, .
\ee
The Hadamard plays an important role in QC by generating the
qubit state from initial spin up or spin down states, i.e.
\be
{\cal H} \mid 0 > = \frac{\mid 0 > + \mid 1>}{\sqrt{2}} \qquad
{\cal H} \mid 1 > = \frac{\mid 0 > - \mid 1>}{\sqrt{2}} .
\ee

Having a Hadamard act in a multi-qubit case, involves operators of
the type $ {\cal  H}\otimes {\bf 1}\otimes{\cal  H},$ for which
Hadamards act on qubits 1 and 3 only. The command for this kind of
operator in \QDENS is \Had[$n_q$,Q] where $n_q$ is the total number of
qubits and the array $Q:{q_1,q_2,...}$  of length $n_q$ indicates which
qubit is or is not acted on by a Hadamard. The rule used is if $ q_i>0$,
then the ith qubit is acted on by a Hadamard, whereas $q_j=0$ designates
that the $j$th qubit is acted on by a unit 2$\times$2 operator. For example,
\Had[3,\{1,0,1\}] has a Hadamard acting on qubits 1 and 3 and a unit 2$\times$2
acts on qubit 2, which is the case given above. To get a Hadamard acting
on all qubits, include all qubits in Q, e.g., use Q=\{1,1,1,....\}. Thus, an
operator \HALL[L]=\Had[L,\{1,1,1....\}]  is also implemented where the array
of 1's has length $n_q$ of all the qubits. Another \QDENS command \had[$n_q$ ,q]
is for a Hadamard acting on one qubit q out of the full set of $n_q$ qubits.
So \QDENS facilitates the  action of a one qubit operator in a multi-qubit environment.

\subsubsection{Usage}

 \QDENS includes several Hadamard commands.  For single qubit cases use either
\(\pmb{ \mathcal{H} }\)
 or {\bf had[1,1]}. For a Hadamard acting on a single
qubit within a set of $L$ qubits use
  \had[L,q];  for a set of Hadamards acting on selected qubits use {\bf \Had[L, \{ $ 0,1,0,1 \cdots$ \}
  ],} and for Hadamards acting on all $L$ qubits use
\HALL[L].  These are demonstrated in the tutorial.

\subsubsection{Rotations}

One can use the rotation operator $R$ to produce a state in any
direction. A rotation {\it about} an axis $\hat{\eta}$ is given in
Eq.~(\ref{eq:rot}).  For special cases,  such as the
$\hat{x},\hat{y},\hat{z}$ and $\gamma=\pi,$ the expanded form
reduce to $-i \sigma_x,-i \sigma_y$ and $-i \sigma_z,$
respectively.  For a general choice of rotation,  one can use the
``MatrixExp" command directly,  or use the spinor rotation matrix
for rotation {\it to} angles $\theta,\phi.$

\begin{figure}[t]
\fbox{\parbox{.8\textwidth}{\includegraphics[width=10cm]{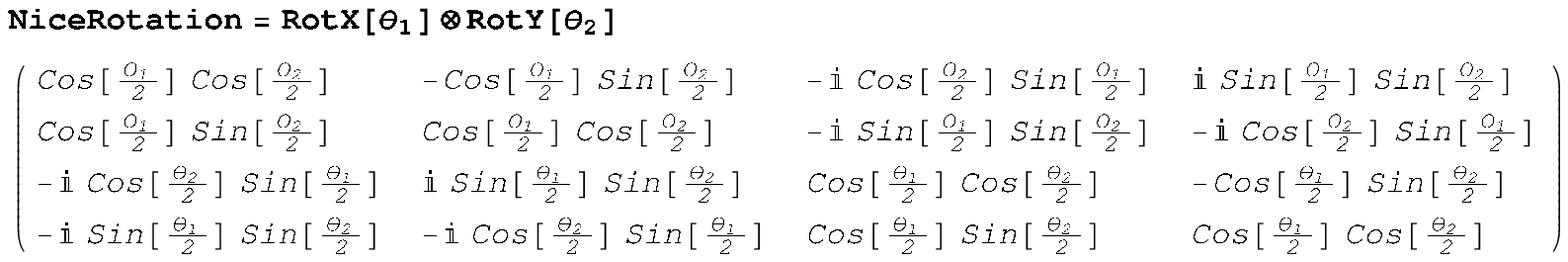}}}
 \caption{Example of multiqubit rotation using \QDENS.}
\protect\label{genrots}
\end{figure}

For a multi-qubit circuit, the rotation operator for say qubits 1
and 3  can be constructed using  the command $R_{\gamma_1}
\otimes {\bf 1}\otimes R_{\gamma_2} \otimes {\bf 1}\otimes \cdots
,$ with associated rotation axes. Examples from \QDENS are given
in Fig.~\ref{genrots}.

\subsubsection{Usage}

Rotation commands for rorarions about the x-, y- or z- axis by an ankle $\theta$
 are included in \QDENS : \RotX[$\theta$], \RotY[$\theta$], \RotZ[$\theta$]  In
 addition,
 {\bf Rotqbit[v,t]}
     builds the matrix corresponding to a rotation around a general
     axis
 axis v by an angle t.

\subsection{Two Qubit Gates}

To produce a quantum computer, which relies on quantum interference,
one must create entangled states.  Thus the basic step of two qubits
interacting must be included.  The interaction of two qubits can take many
 forms depending on the associated underlying dynamics.  It is helpful in QC,
   to isolate certain classes of interactions that can be used as logical
    gates within a circuit model.

\subsubsection{CNOT}

The most commonly used two-qubit gate is the controlled-not (\cnot) gate.
The logic of this gate is summarized by the expression
 $$ \cnot \mid c, t > = \mid c,  t \oplus c>,$$ where
$c=0,1$  is the control bit and $t=0,1$ is the target bit. In a
circuit diagram the $\bullet$ indicates the control qubits and the
$\oplus$ indicates the target qubit.

\begin{equation}
\Qcircuit @C=1em @R=0.5em @!R {
 &\qw    &\ctrl{1}& \qw  & \qw \\
 &\qw    &\targ   &\qw   & \qw
}
\nonumber
\end{equation}

The
final state of the target is denoted as ``$t \oplus c$" where
$\oplus$ addition is understood to be modular base 2.  Thus, the
gate has the logical role of the following changes (control bit
first) $ \mid 0 0>{\small \mapsto} \mid 0 0>;   \mid 0 1>{\small \mapsto}
\mid 0 1>;\nobreak{ \mid 1 0>{\small \mapsto}\mid 0 1>;}
 \nobreak{ \mid 1 1>{\small \mapsto} \mid 1 0> .}$   All of this can be simply stated
 using projection and spin operators as
\be
\cnot[ c , t] =  \mid 0>_c < 0\mid\otimes {\bf I }_t  +   \mid 1>_c
< 1\mid \otimes  \sigma^t_x , \ee  with $c$ and $t$ denoting the
control and target qubit spaces.  The \cnot, which is briefly expressed
 as $\cnot =  {\cal P}_0 {\bf I }   + {\cal P}_1   \sigma_x ,$
 is  called the controlled-not gate since \NOT $\equiv \sigma_x.$
  A matrix form for this operator acting in a two qubit space is
\be
\cnot =
 \left(
\begin{array}{rrrrrrrrrr}
1 &&& 0 &&& 0 &&& 0\\
0 &&& 1 &&& 0 &&& 0\\
0 &&& 0 &&& 0 &&& 1\\
0 &&& 0 &&& 1 &&& 0\\
\end{array}\right) .
\label{cnot}
\ee The rows \& columns  are ordered numerically as:\ $
00,01,10,11.$

The \cnot gate is used extensively in QC in a multi-qubit context.
Therefore, \QDENS gives a direct way to embed a two-qubit \cnot
into a multi-qubit environment. The command is
\cnot$[n_q,c,t]$ where $n_q$  is the total number of qubits and $c$ and
$t$ are the control and target qubits respectively as in the following
examples: If the number of qubits is 6, and a \cnot acts with
qubit 3 as the control and 5 is the target, the operator (which is a $2^6\times2^6$ matrix)
is invoked by the command $\cnot[6,3,5].$ The command $\cnot[6,5,3]$ has 6 qubits,
with qubit 5 the control and 3 the target. The basic case in Eq.(\ref{cnot}) is therefore just $\cnot[2,1,2].$

\subsubsection{CPHASE}

Other two qubit operators can now be readily generated. The \cphase gate, which plays
an important role in the cluster model of QC is simply a controlled $\sigma_z$
\be
\cphase[ c , t] =  \mid 0>_c < 0\mid\otimes {\bf I }_t  +   \mid 1>_c < 1\mid \otimes  \sigma^t_z ,
\ee
which in  two qubit space has the matrix form
\be
\cphase =
 \left(
\begin{array}{rrrrrrrrrr}
1&&& 0&&& 0&&&0\\
0&&& 1&&& 0&&&0\\
0&&& 0&&& 1&&&0\\
0&&& 0&&& 0&&&-1\\
\end{array}\right) .
\ee
The multi-qubit version is $\cphase[n_q, c, t],$ with the same rules as for the $\cnot$ gate.

\subsubsection{Other Gates}

The generation of other gates, such as swap gates~\footnote{See the \QDENS command \Swap.},
Controlled-$i \sigma_y$ ( also known as a \crot gate) are now clearly extensions of the
prior discussion.

The swap gate swaps the content of two qubits. It can be decomposed in a
chain of \cnot gates:

\begin{equation}
\Qcircuit @C=1em @R=0.5em @!R {
\lstick{|\Psi_1\rangle}   &\qw    &\ctrl{1}& \qw  & \qw & \targ
  &\qw& \qw & \ctrl{1} &\qw&\rstick{|\Psi_2\rangle}  \\
\lstick{|\Psi_2\rangle}   &\qw    &\targ   &\qw   & \qw & \ctrl{-1}
 &\qw& \qw & \targ    &\qw&\rstick{|\Psi_1\rangle}
}
\end{equation}

Another example is $\crot$
\be
\crot[ c , t] =  \mid 0>_c < 0\mid\otimes {\bf I }_t  +   \mid 1>_c < 1\mid
 \otimes  i \, \sigma^t_y ,
\ee
which, in two qubit space, has the matrix form
\be
\crot =
 \left(
\begin{array}{rrrrrrrrrr}
1 &&& 0 &&& 0 &&& 0\\
0 &&& 1 &&& 0 &&& 0\\
0 &&& 0 &&& 0 &&&1\\
0 &&& 0 &&& -1 &&& 0\\
\end{array}\right) .
\ee
A multi-qubit version of $\crot[n_q, c, t],$ with the same rules as for
the $\cnot$ gate can easily be generated by a modification of {\it Qdensity.m}.

Indeed, the general case of a controlled-$\Omega,$ where $\Omega$ is any
one-qubit operator is now clearly
\be
C\Omega[ c , t] =  \mid 0>_c < 0\mid\otimes {\bf I }_t  +   \mid 1>_c < 1\mid
 \otimes  \, \Omega^t ,
\ee
with corresponding extensions to  matrix and multi-qubit
renditions.

\subsubsection{Usage}

\QDENS includes the following two-qubit operators, acting between
qubits c(control) and t(target) imbedded in a system of $L$ qubits:
\cnot[L,c,t], \cphase[L,c,t],  \ControlledX[L,c,t],
\ControlledY[L,c,t] and   \Swap[L,$q_1$,$q_2$].   A generic two
qubit operator within a multiqubit system involving operators $Op1 $
and $Op2 $ is {\bf TwoOp}[L,$q_1$,$q_2$,Op1,Op2].

\subsection{Three Qubit Gates}

The above procedure can be generalized to three qubit operators.
The most important three qubit gate is the Toffoli~\cite{Nielsen}gate,  which
has two control bits that determine if a unit or a NOT($\sigma_x$)
operator acts on the third (target) bit. The projection operator
version of the Toffoli is simply
\be
{\rm \Toffoli } \equiv{\cal P}_0 \otimes {\cal P}_0 \otimes {\bf 1}
+  {\cal P}_0 \otimes {\cal P}_1 \otimes {\bf 1}
+ {\cal P}_1 \otimes {\cal P}_0 \otimes {\bf 1}
+ {\cal P}_1 \otimes {\cal P}_1 \otimes \sigma_x,
\ee
which states that the third qubit is flipped only if the first two(control)
qubits are both 1.

For a multi-qubit system
the \QDENS command \Toffoli[$n_q$ ,$q_1$, $q_2$, $q_3$] returns the Toffoli
operator with $q_1$ and $q_2$ as control qubits, and $q_3$ as the target
qubit within the full set of $n_q$ qubits.

The Toffoli gate can be specialized or reduced to lower gates and is a
universal gate.

\subsubsection{Usage}

\QDENS includes  a generic three qubit operator within a multiqubit
system involving operators $Op1,$  $Op2, $ and $Op3 $ : {\bf
ThreeOp}[L,$q_1$,$q_2$,$q_3$,Op1,Op2,Op3].  The Toffoli gate is a
special case and is invoked by the command \Toffoli[L,c,c,t],  where
$c,c,t$ specifies the two control and the one target qubit out of
the set of $L$ qubits.

\section{SPECIAL STATES}
\label{sec6}

As a prelude to discussing QC algorithms, it is useful to examine
how to produce several states that are part of the initialization
of a quantum computer. These are the uniform superposition, the
two-qubit Bell,~\cite{Bell}  three-qubit GHZ,~\cite{GHZ} and
Werner~\cite{Werner} states.

\subsection{Uniform superposition}

In many QC processes, the first step is to produce an initial state of
the $n_q$ qubits that is a uniform superposition of all of its possible
computational basis states. It is the initialization of this superposition
that allows a QC to address lots of questions simultaneously and is
often referred to as the ``massively parallel" feature of a quantum computer.

The steps start with a state of all spin-up $ \mid 0 0 0 0 \cdots
>$, then every qubit is acted on by a Hadamard ${\cal
H}\otimes{\cal H}\otimes{\cal H}\otimes \dots,$ which is done by
the \QDENS command  \HALL[$n_q$]. Thus each up state is replaced
by ${\cal H} \mid 0> = \frac{\mid 0> + \mid 1 >}{\sqrt{2}}$, and
we have the uniform superposition \be \mid \Psi >={\rm HALL}[n_q]
\mid 0> = \frac{1}{2^{n_q/2}} \sum_{x=0}^{2^{n_q}-1} \mid x > ,
\ee where $x$ is the decimal equivalent to all possible binary
numbers of length $n_q$.

\begin{figure}[t]
\fbox{\parbox{.8\textwidth}{\includegraphics[width=8cm]{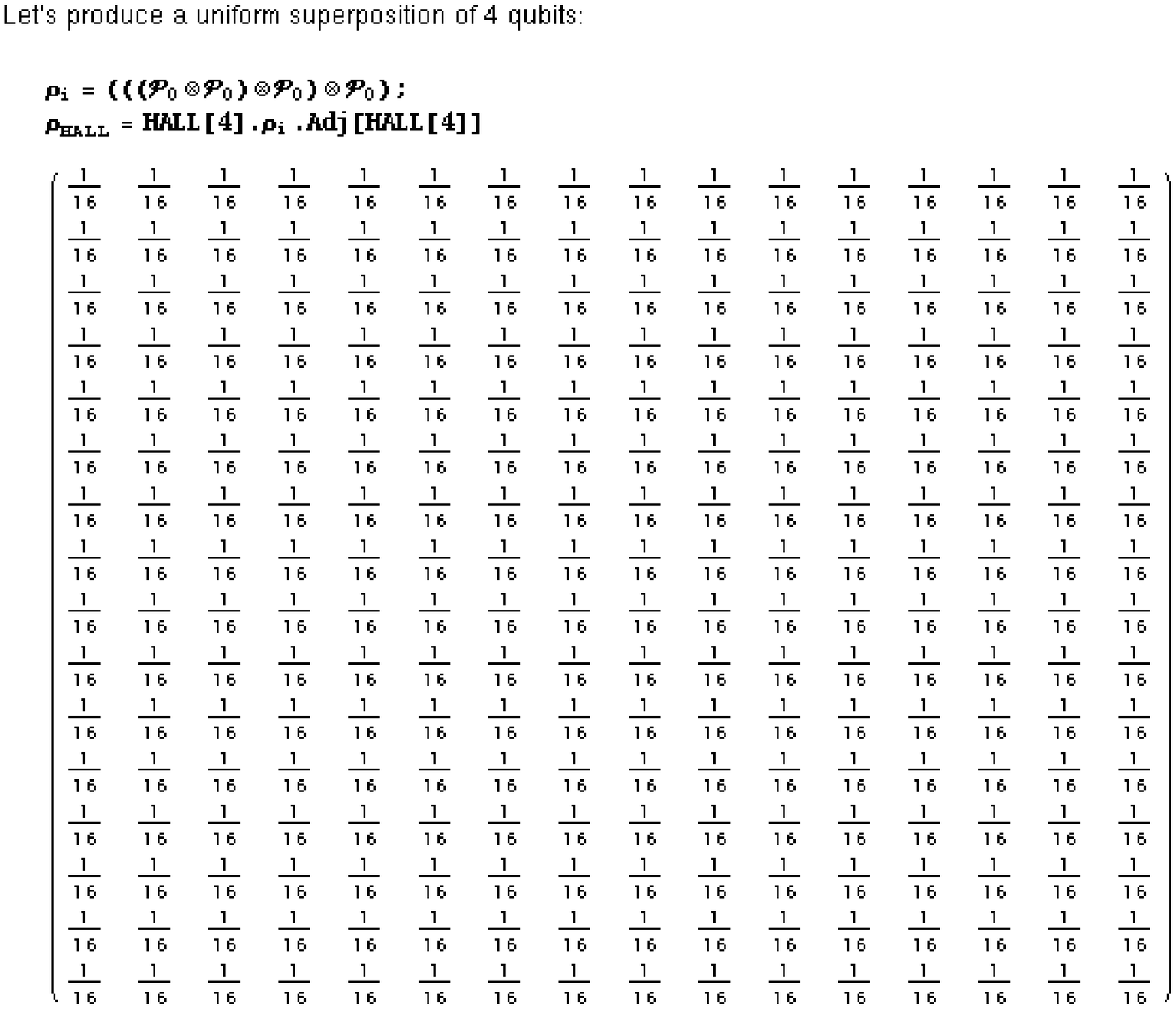}}}
 \caption{Construction of a uniform four qubit state.}
\protect\label{uniform}
\end{figure}

An example of this process, including the associated density matrices,
is in Fig.~\ref{uniform}.

\subsection{Bell States}

The singlet and triplet states are familiar in QM as the total spin zero and
one states, with zero spin projection ($M=0$). They are the basic states that
enter into the EPR discussion and they are characterized by their ``entanglement."
Bell introduced two more combinations, so that for two spin 1/2 systems the
four Bell states are:
\bea
\mid B_{ 0 0 }>&=&  \frac{1}{\sqrt{2}}\, \mid 0 0 > + \mid 1 1 > \nonumber \\
\mid B_{ 0 1 }>&=&  \frac{1}{\sqrt{2}}\, \mid 0 1 > + \mid 1 0 > \nonumber \\
\mid B_{ 1 0 }>&=&  \frac{1}{\sqrt{2}}\, \mid 0 0 > - \mid 1 1 > \nonumber \\
\mid B_{ 1 1 }>&=&  \frac{1}{\sqrt{2}}\, \mid 0 1 > - \mid 1 0 > ,
\eea
or in one line $ \mid B_{ a b }> =  \frac{1}{\sqrt{2}}\, \mid 0 b >
 + (-1)^a \mid 1 \bar{b} >$,
where $\bar{q}$ is the NOT[q] operation.

A circuit that produces these states, starting from the state $\mid a b>$ ($ a,b = 1,0$)
consists of a Hadamard on qubit one, followed by a \cnot.
The \QDENS version is thus:\\
B [a\_, b\_] :=\cnot[2,1,2]\Had[2,\{1,0\}](Ket[a]$\otimes$Ket[b]).

\begin{figure}[t]
\fbox{\parbox{.8\textwidth}{\includegraphics[width=10cm]{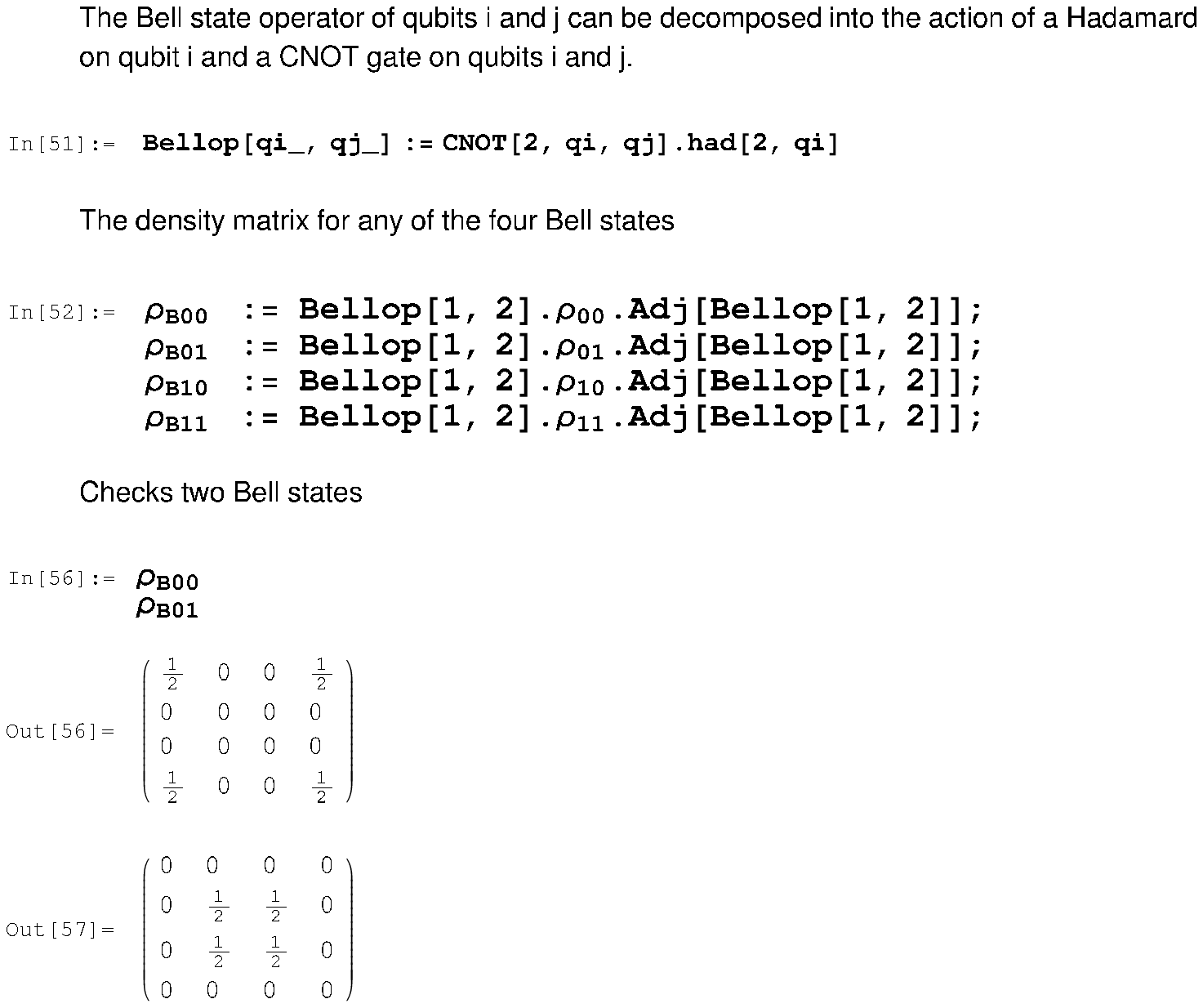}}}
 \caption{Example from {\it Teleportation.nb}}
\protect\label{Bell}
\end{figure}

The density matrix version involves defining the unitary
transformation $U\equiv \cnot[2, 1,2] . {\rm Had}[2, {1,0}]$ and
an initial density matrix $\rho^I_{a b} \equiv \mid a b > < a b
\mid$, then evolving to the density matrix for each of the Bell
states
\be
\rho^{Bell}_{ a b } = U \cdot \rho^I_{a b} \cdot U^\dagger.
\ee
In Fig.~\ref{Bell} part of this process taken from {\it Tutorial.nb}
is shown. The tutorial includes a demonstration that the Bell states have zero
polarization  (as is obvious from their
definition), and a simple diagonal form for the associated tensor
polarization $\overleftrightarrow{T}.$

Another useful application shown in the tutorial is that taking
the partial traces of the Bell state density matrices, yield non
pure single qubit density matrices and that the associated von
Neumann entropy defined by $\nobreak{S[\rho] = - {\rm Tr}[\rho
\;\log_2 \rho]}$ is zero for the Bell states, but 1 for the single
qubit density matrices $\rho_1$ and $\rho_2$. Thus each qubit is
in a more chaotic state, which physically means they have zero
average polarization.~\footnote{Since many different state vectors
can yield a net zero average polarization, it is clear  that
the density matrix stores less information than in a state vextor, albeit
 realistic statistical,
information.} This property is an indication of the entanglement
of the Bell states.

\subsection{GHZ States}

Three qubit states that are similar in spirit to the Bell states,
were introduced by Greenberger, M. A. Horne, and A.
Zeilinger~\cite{GHZ}. The basic GHZ state is  \be \mid \Psi> =
\mid {\rm GHZ}> = \frac{1}{\sqrt{2}} \, ( \, \mid 0 0 0> + \mid 1
1 1>), \ee which may be written: \be \mid GHZ>=U_{GHZ} \mid 000>
\ee with $U_{GHZ}={\rm CNOT}[3,1,2].{\rm CNOT[3,1,3]}.{\rm
had[3,1]}$ which corresponds to the following circuit:
\begin{equation}
\Qcircuit @C=1em @R=0.5em @!R {
\lstick{|0\rangle} &\qw    &\gate{H} &\qw   & \ctrl{1}  & \qw & \ctrl{2} &\qw \\
\lstick{|0\rangle} &\qw    &\qw      &\qw   & \targ     & \qw & \qw      &\qw\\
\lstick{|0\rangle} &\qw    &\qw      &\qw   & \qw       & \qw &  \targ   &\qw
}
\nonumber
\end{equation}

 A complete set of  eight GHZ states can be produced by the step \be
U_{\rm GHZ} \mid abc> = \mid {\rm GHZ}_{abc}> = \frac{1}{\sqrt{2}}
( \mid 0 b c > + (-1)^a \mid 1 \bar{b} \bar{c}> ) . \ee   For
 all eight of these three qubit states,  the associated density matrix can be formed
   $\rho_{1 2 3}^{abc}=\mid {\rm GHZ}_{abc}><
{\rm GHZ}_{abc} \mid$ and  are seen in {\it Tutorial.nb} to have a simple
 structure.  Taking partial traces to generate the two qubit
$\rho_{ 1 2},\rho_{ 1 3},\rho_{ 2 3}$ and single qubit $\rho_{ 1
},\rho_{ 2 },\rho_{ 3 }$ density matrices,  we see that for these
states every qubit has zero polarization and a simple structure
for the pair and three qubit correlation functions.  In addition,
the entropy of these GHZ set of states is zero and the
sub-entropies of the qubit pairs and single qubits are all 1,
corresponding to maximum disorder.

\begin{figure}[t]
\fbox{\parbox{.8\textwidth}{\includegraphics[width=10cm]{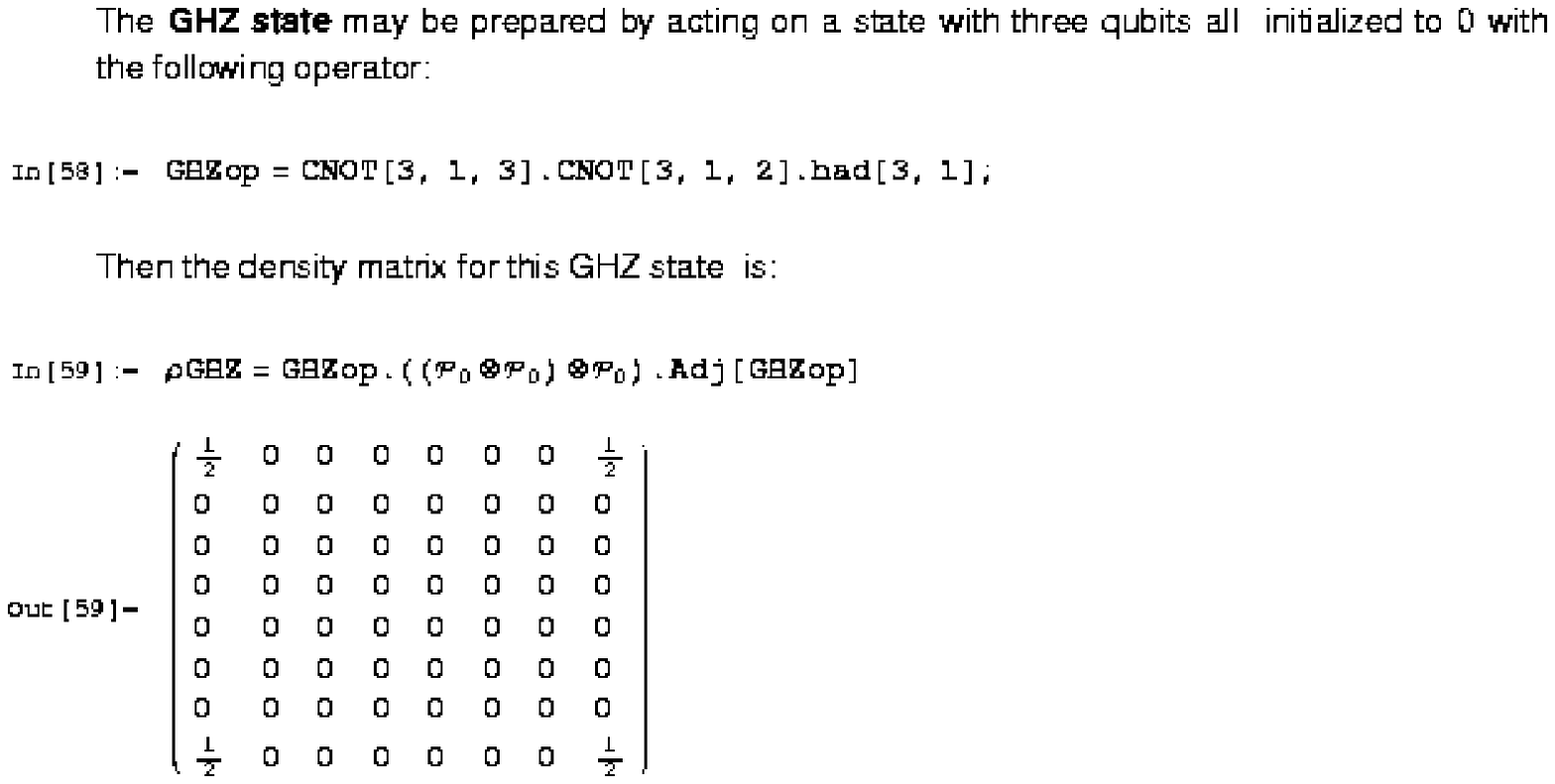}}}
 \caption{Construction of a GHZ state. Example from {\it Tutorial.nb}}
\protect\label{GHZ}
\end{figure}

A sample GHZ realization in \QDENS is given in Fig.~\ref{GHZ}

\subsection{Werner States}

Another set of states were proposed by Werner~\cite{Werner}.  They are defined
in terms of a density matrix: \be \rho_W  =  \lambda \rho_B + (1-
\lambda) \rho_u \otimes \rho_u,\ee where $0 \leq \lambda \leq 1 $
is a real parameter. The limit $\lambda=1$ yields a completely
entangled Bell state density matrix $\rho_B = \mid B_{a b} ><B_{a
b} \mid,$ whereas lower values of $\lambda$ reduce the
entanglement.  The $\lambda=0$ limit gives a simple two qubit
product $\rho_u \otimes \rho_u,$ where $\rho_u=\frac{ {\bf 1}}{2}$
is the density matrix for a single qubit with zero polarization,
i.e. it corresponds to a chaotic limit for each qubit.  Therefore,
the parameter $\lambda$ can alter the original fully entangled
Bell state by introducing chaos or noise.  Therefore, the Werner
state is called a state of noisy entanglement.

 The entropy of a two qubit
Werner state as a function of $\lambda$ ranges from
two~\footnote{This corresponds to an entropy per qubit of 1.} for
$\lambda=0 ,$  to zero for $\lambda=1.$ The entropy of the single
qubit state is 1.  See {\it Tutorial.nb} for a sample Werner  \QDENS realization.

\section{TELEPORTATION}
\label{sec7}
To understand QC teleportation, let us first consider classical
teleportation, which entails only classical laws. For example,
suppose Alice measures a vase using a laser beam to measure its
dimensions, shape, color and decoration. She therefore has a full
binary description of the vase in 3-D. Bob has all the material to
make another vase and, upon receiving the file that Alice sends
him by computer, is able to make an exact copy of the vase. There
are only local operations (LO) (measuring and sending by Alice and
reconstruction by Bob) and a classical communication (CC); so this
called a LOCC process. How does this differ from teleportation
using Quantum Physics?

In Quantum Mechanics, measurement affects the sample; as a result,
after collecting the information to send to Bob, the original
sample is no longer in its original state. In the classical case,
one ends up with two identical copies. In the QC case, Bob has the
only extant system. Another difference is that in the QC case,
Alice and Bob share an entangled state, say an EPR or a Bell
state, Alice entangles the original system with one member of the
pair, then measures and by LOCC sends Bob her result. By virtue of
the shared entanglement, information is shared by the LOCC and by
the Quantum effect of sharing the entangled state. Some
information is transmitted by a ``Quantum channel.''  Therefore,
the information sent by computer is less than needed in the
classical case, because it is supplemented by the Quantum transfer
of entanglement. The strange nature of Quantum transportation is
thus no stranger than the EPR/Bell effect, which has been affirmed
experimentally.

To understand these general remarks, let us use \QDENS to examine three cases.

\subsection{One Qubit Teleportation}

\begin{figure}[tbh]
\vspace{9pt}
\includegraphics[]{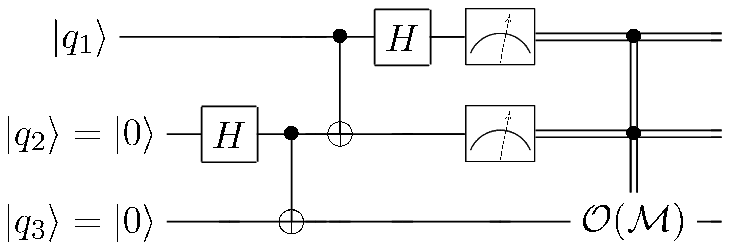}
\caption{Schematic circuit for one qubit quantum teleportation.}
\label{teleport1}
\end{figure}

Suppose Alice has one qubit $q_1$ in an unknown state $\mid \Psi>=
a_0  \mid 0> + a_1 \mid 1>,$ with an associated spin density
matrix $\rho_0 \equiv \mid \Psi><\Psi \mid$. In \QDENS, such a
state is generated randomly. Bob and Alice share a two qubit
entangled state, which we take as one of the Bell states $B_{q_2,
q_3}$. This Bell state could be provided by an outside EPR
purveyor, but for convenience let us assume that Bob produces the
entangled pair and sends one member of the pair $\mid q_2>$ to
Alice, as shown in Fig.~\ref{teleport1}. Alice then entangles her
state $\mid \Psi>,$ using the inverse of the steps that Bob
employed to produce the entangled pair, and then she measures the
state of her $\mid \Psi> \otimes \mid q_2>,$ which yields a single
number between zero and three (or one of the binary pairs $  0 0;
0 1; 1 0; 1 1).$ Alice transmits by CC that number to Bob, who
then knows what to do to his qubit $q_3$ in order to slip it over
to the state $\mid \Psi >,$ without making any measurements on it
to avoid its destruction. In the end, Alice no longer has a qubit
in the original state, but by LOCC and shared entanglement, she is
happy to know that Bob has such a qubit, albeit not made of the
original material. The only material transmitted is the single
member of the entangled pair.

In \QDENS, the following detailed steps are shown in the file {\it Teleportation.nb}:
construction of an initial random qubit for Alice, Bob's entanglement process,
Alice's entanglement, measurement and CC and Bob's consequent actions. This process
is described using the density matrix language.  A rendition using the
quantum states directly is easily generated.

\subsection{Two Qubit Teleportation}
\subsubsection{ Two EPR Teleportation}

\begin{figure}[htb]
\vspace{9pt}
\includegraphics[]{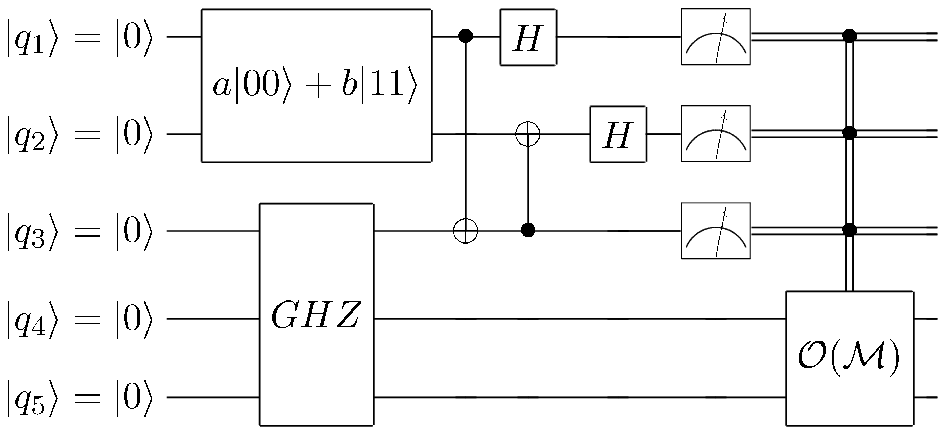}
\caption{EPR teleportation using a GHZ entangled state.}
\label{teleportEPR}
\end{figure}

A similar process can be invoked to teleport an initially unknown
state of two qubits $\mid \Psi> =\sum_{i,j=0,1}\  a_{ i j }
\mid q_{1i} q_{2j} >$. In the special case that the unknown state is one
of the Bell states, it can be transported using the procedure
shown in Fig.~\ref{teleportEPR}. Bob now prepares a three qubit
entangled GHZ state using Hadamards and \cnot \!\!s and then sends one
of the three qubits $q_3$ over to Alice, who entangles that one
with her original $q_1,q_2$ qubits $\mid \Psi>_{12}\otimes \mid
q_3>$ and then measures the state of her three qubits, with the
result of a number between zero and seven ( e.g. one of the binary
results $000;001;010,011;100;101;110;111$). She transmits that
decimal number to Bob by CC ( a phone call say), who then knows
what to do to put his two qubits $q_4 ,q_5$ into the original
state $\mid \Psi>.$ Again all the steps are presented in detail in
\QDENS in the file {\it Teleportation.nb}.

\subsubsection{ General Two Qubit Teleportation}

\begin{figure}[tbh]
\vspace{9pt}
\includegraphics[]{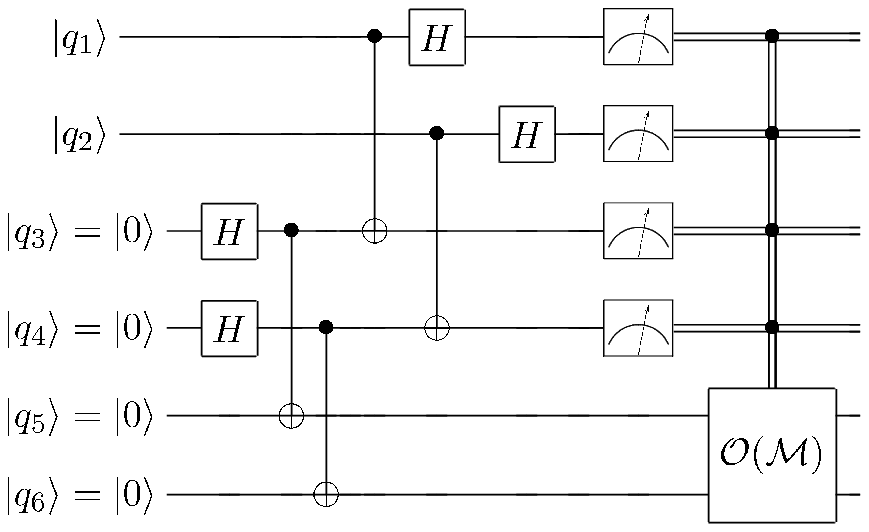}
\caption{Schematic circuit for two qubit quantum teleportation.}
\label{teleportGen2}
\end{figure}

The two qubits can be in a more general state than in the above
discussion which was restricted to being one of the Bell states. In
this case, Bob needs to entangle four qubits by a chain of Hadamard and
\cnot gates as shown in Fig.~\ref{teleportGen2}. He then sends two qubits
$q_3,q_4$ over to Alice who entangles her two with them. That is, she
entangles $\mid \Psi>_{1 2} \otimes \mid q_3>\otimes \mid q_4>,$ then measures
them with a decimal number result that is between zero and 14 or a binary
measurement of: $ 0000; 0001;0010;0011; \cdots ;1111$. With that number,
Bob knows what to do and places his two qubits into the original state
$\mid \Psi>$. Again see {\it Teleportation.nb} for the detailed layout.

\begin{figure}[t!]
\vspace{9pt}
\fbox{\parbox{1.\textwidth}{\includegraphics[width=10cm]{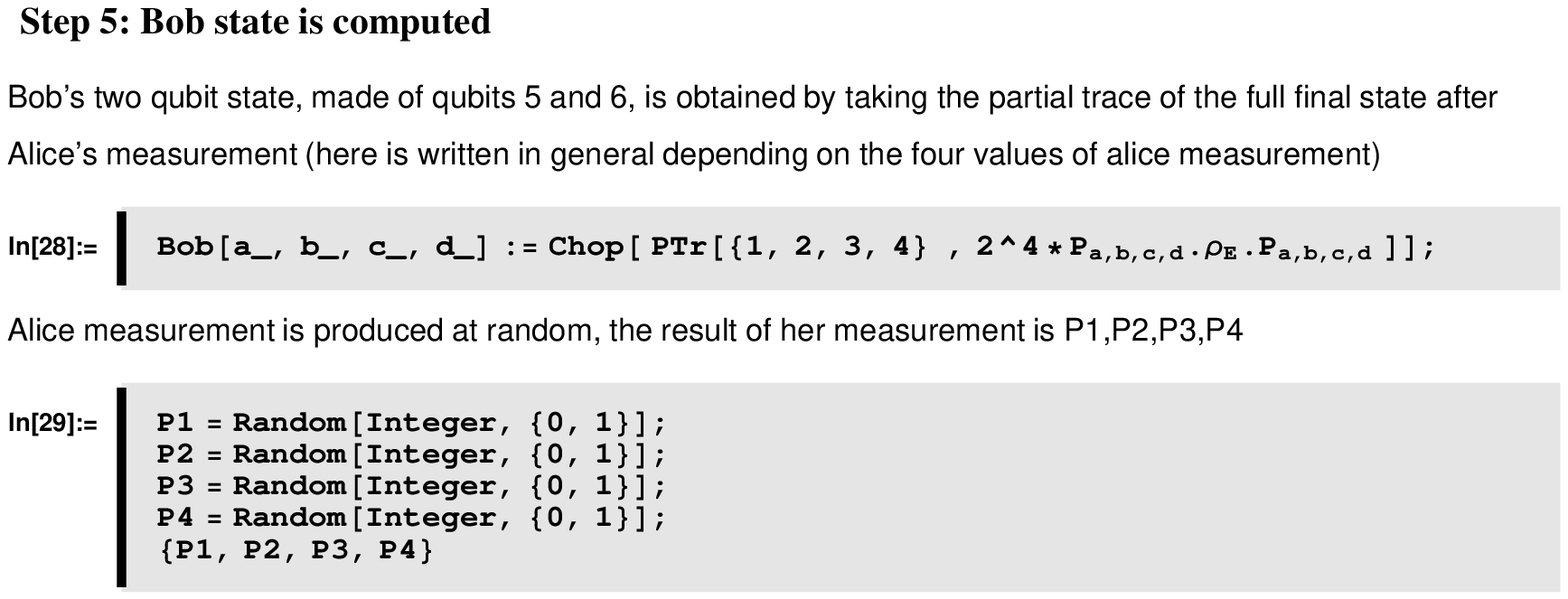}}}
\caption{Example from {\it Teleportation.nb}}
\label{fig:tele}
\end{figure}

\section{GROVER'S SEARCH}
\label{sec8}

Assume you have a set of items and you want to find a particular
one in the set which has been marked beforehand. Let us further
restrict the problem by saying that you are only allowed to ask
yes/no questions, e.g. ``Is this item the item?'' In a disordered database
with N items and one marked item that problem would require on the
order of N trials to find the marked item. Quantum mechanics allows
the states of a system to exist in superpositions and thus in many
cases permits one to parallelize the process in some sense.
Grover~\cite{Grover} proposed an algorithm that lowers the number of
trials needed to $O(\sqrt{N})$ by making clever use of interference and
superposition. He based his ideas on an analogy to the multiple slit
problem~\cite{Groverslit}.

\subsection{The Oracle}

Grover's search algorithm relies on the use of an Oracle. The idea
underlying the Oracle is the following: the Oracle is a function
that can recognize a solution to a problem although it may not
know how to solve the problem. In that sense a classical example
could be a lock and a key, the problem of finding the proper key that
would open a lock out of a bunch of keys illustrates the role of the lock as
an Oracle: you select a key and you try it, the Oracle would tell you
whether that was the correct key or not, but you cannot ask the lock to
single out the key from the bunch. The essential difference between
a classical and a quantum Oracle is that in the quantum case the Oracle
can act on an input which is a superposition of all states. In our example
that would mean that we can try a superposition of all the keys at the
same time.

The above description of the role of an Oracle takes on the explicit form
\be
{\bf ORACLE} \mid x>_N  \mid y>_1 = \mid x >_N  \mid y \oplus f(x) >_1,
\ee  which involves an $N$ qubit and a single qubit product space and a specified function $f(x).$  The matrix form of the Oracle is
\be
< x' \mid < y'  \mid {\bf ORACLE} \mid x>   \mid y>  = \delta_{x' x}  <y'\mid y \oplus f(x) >.
\ee  Examples of this Oracle matrix for single and double marked items are given in detail in
the Grover.nb  notebook,  where the ``inversion about the mean"  process is also presented and explained in detail.

\subsection{One marked item}

A schematic description of the searching process is:

\begin{equation}
\Qcircuit @C=0.5em @R=0.2em @!R {
&&&&&\multigate{2}{H^{\otimes n}}  &\qw& \qw & \multigate{6}{G}  &\qw  &\qw  &\multigate{6}{G}& \qw &\hdots &&\qw  &\qw  &\multigate{6}{G} & \qw &\qw \\
&&&\lstick{|0\rangle}& &\ghost{H^{\otimes n}} &\qw  &\qw   &\ghost{G} &\qw  &\qw  &\ghost{G} & \qw &\hdots &&\qw  &\qw  &\ghost{G} & \qw &\qw &&\rstick{\rm Out\; register}\\
&&&  &   &\ghost{H^{\otimes n}}   &\qw   &\qw  &\ghost{G}  &\qw  &\qw  &\ghost{G}& \qw &\hdots &&\qw  &\qw  &\ghost{G} & \qw & \qw \\
\vdots\\
            &&&& &\qw    &\qw  &\qw  &\ghost{G}  &\qw  &\qw  &\ghost{G} & \qw &\hdots &&\qw  &\qw  &\ghost{G} & \qw &\qw\\
{\rm Oracle}&&&& &\qw    &\qw   &\qw &\ghost{G}  &\qw  &\qw  &\ghost{G} & \qw &\hdots &&\qw  &\qw  &\ghost{G} & \qw&\qw \\
{\rm space} &&&& &\qw    &\qw   &\qw  &\ghost{G}  &\qw  &\qw  &\ghost{G} & \qw &\hdots &&\qw  &\qw  &\ghost{G} & \qw&\qw \\
}
\nonumber
\end{equation}

The basic steps, which are detailed in {\it Grover.nb}, consist of
applying a Hadamard to all qubits in the register while setting the
Oracle to recognize the marked item. Then we need to construct
the Grover operator and apply it a number of times. Finally we measure
the output register which then will give the marked item with high
probability. The probability of success depends on the number of
times the Grover operator has been applied, as can be studied in
the notebook.

{\it Grover.nb} contains two examples, the first one has only one
marked item in the full database. The size of the database, given
by the number of qubits, can be varied, together with the number of
times the Grover operator is applied.

\subsection{Two marked items}

The second example includes two marked items in the database (and may
be generalized to the case of M marked items). Of course, one
needs to enlarge the register and Oracle space.

\section{SHOR'S ALGORITHM}
\label{sec9}

Shor's factoring algorithm is the most celebrated quantum
algorithm, partly due to its powerful use of quantum superposition
and interference to tackle one of the problems upon which most of
our secure bank transactions rely. The factoring algorithm,
described in detail in the notebook {\it Shor.nb}, essentially
first relates the problem of factoring a number, N, made up of two
prime numbers, N=$p\times q$, to the problem of finding the period
of a particular function $x^j \mod(N)$, being $x$ a
coprime\footnote{Two positive integers $a$ and $b$ are said to be
coprime if they have no common factor other than 1, that is, their
greatest common divisor is 1.} to N smaller than N. Then finding
the period is related to the computation of the Quantum Fourier
Transform (QFT) (analog to the discrete Fourier transform), for
which a very effective quantum algorithm exists.

Schematically, the procedure is the following: build two registers, 1 and 2,
with $n_1$ and $n_2$ qubits each, initially set to $|0>$. Then Hadamard the first
register, the state of the full system then reads:
\be
\Psi = {1\over 2^{n_1}} \sum_{i=0}^{2^{n_1}-1} \mid i>_1 \otimes \mid 0>_2 \, .
\ee

Then we take a number $x$ smaller than N and coprime with it and load the second
register with the function $x^i \mod(N)$, giving:

\be
\Psi = {1\over 2^{n_1}} \sum_{i=0}^{2^{n_1}-1} \mid i>_1 \otimes \mid x^i \mod N>_2 \, .
\ee

At this point, a measurement is performed on the second register and then one
applies the QFT to the first register. From that value measured in the first
register, one is able, with a certain probability, to factor the original
number N.

A detailed study of the performance of the algorithm, e.g. analysis of
probabilities of success depending on the size of register considered
experimentally can be done within the notebook. A thorough theoretical
description of the algorithm can be found in Refs.~\cite{Shor,Gerjuoy}.

{\it Shor.nb} contains four slightly different approaches to the
algorithm, mainly differing in the procedure used to compute the
QFT. The most constructive case and also the one appropriate when
studying possible noise effects on parts of the circuit, is the
density matrix one. There the QFT (see also {\it QFT.nb}) is
obtained using unitary operators in the same way as occurs
experimentally. That full QFT treatment implies that the QM method
is not practicable for heavy computing simulations, say when
number of qubits goes above 10.

Then there are three other cases, two of them using the state
vector expression for the QFT with and without explicit
construction of both registers.

\begin{figure}[t]
\fbox{\parbox{1.\textwidth}{\includegraphics[width=10cm]{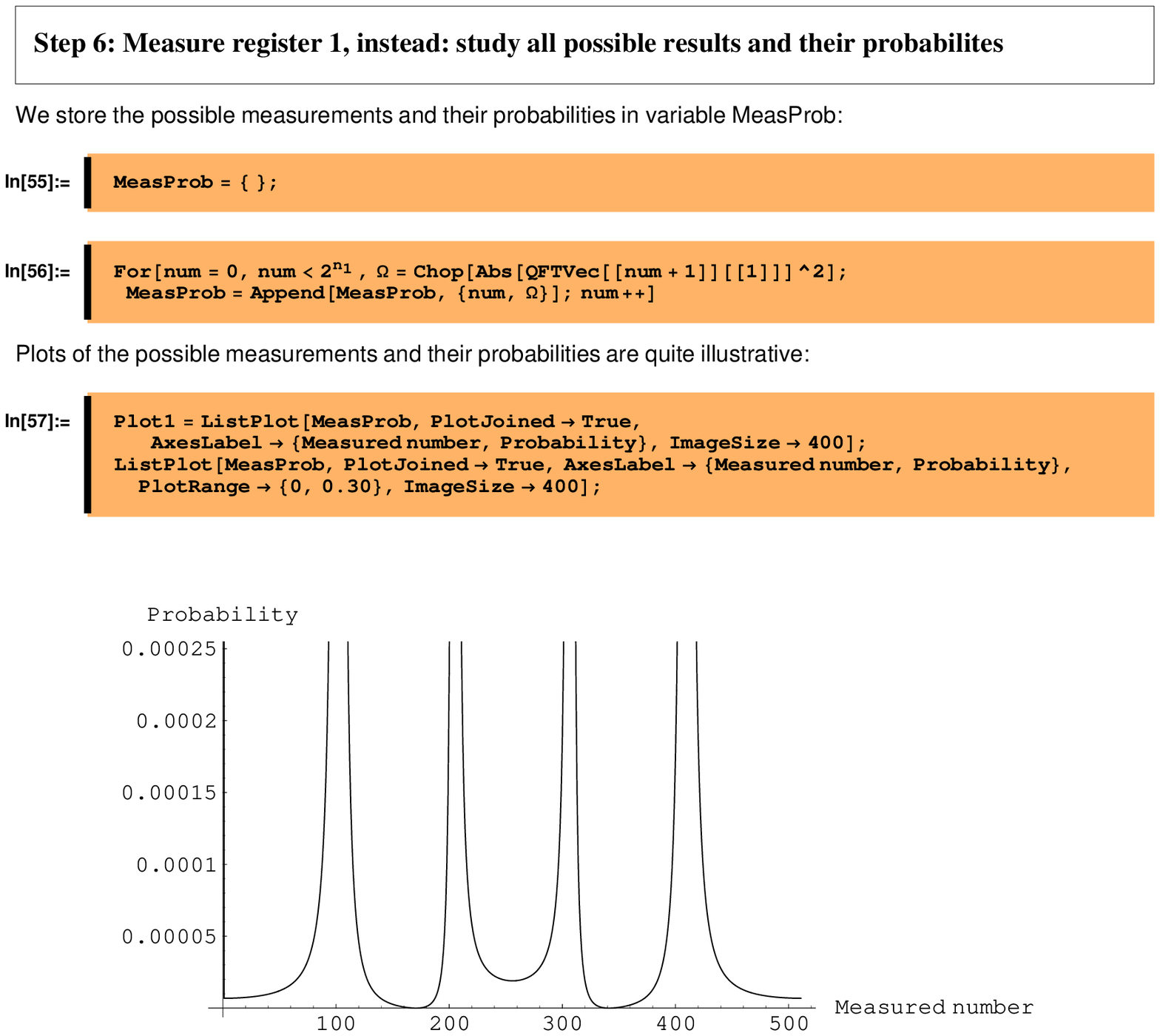}}}
 \caption{Example from {\it Shor.nb}}
\protect\label{fig:shornb}
\end{figure}

Finally the last example, makes use of the already implemented
discrete Fourier transform in Mathematica. This example is thus
useful when  emphasis is  on studying larger numbers to check
probabilities of success, without concern for the actual quantum
mechanical way of building the QFT. In Fig.~\ref{fig:shornb} a
snapshot from {\it Shor.nb} is shown.

\section{CLUSTER MODEL}
\label{sec10}

An alternative to the circuit model for QC has been suggested in a
series of recent papers~\cite{Cluster}. The basic idea of this
approach is to start with an initial state that is highly
entangled by virtue of two-qubit \cphase operations between
nearest neighbors in a cluster of qubits. The \cphase operations
could be generated somehow by Ising model spin-spin interactions.
Once an appropriate cluster is designed, then a carefully planned
set of single qubit measurements are made in various directions.
The results of those measurements are passed on by classical
communications, until one reaches a final qubit, or set of qubits,
from which a result can be deduced once a local correction
involving Pauli operators and the binary results of the
measurement is invoked. This method is being developed, with the
procedures for general algorithms still being formulated. It is
however novel and promising, especially since it involves single
qubit measurements, can generate gates without use of magnetic
fields to rotate spins, and holds forth the promise of error
stability. It does require however a large increase in the number
of qubits. The fact that measurement collapse out qubits and
essentially destroys the initial state is why this approach is
called ``one-way computing." Of course, one can reconstruct the
initial state and try again, which is typical of QC. One can also
reuse qubits after measurement.

It turns out that the modular nature of \QDENS allows one to use it
to demonstrate and test some of the basic ideas of the cluster model
of QC. The first illustration is the basic transport process involving
two qubits. The two qubits are initially in a $\mid \Psi>\otimes \mid + >$ state,
where $\mid \Psi>$ is a general unknown one qubit state. Then a \cphase
operator acts between the two qubits, followed by a measurement of qubit
one in the $\mid +x> \equiv \mid +>$ direction with a result of either
$a=$ zero or one. The second qubit proves to be in the state
$ \sigma_x^a \mid \Psi>$. It is simple to confirm this algebraically. It
forms the basic building block of the cluster model. It is illustrated
in \QDENS using a density matrix example.

\begin{figure}[t]
\centering
\includegraphics[width=5cm]{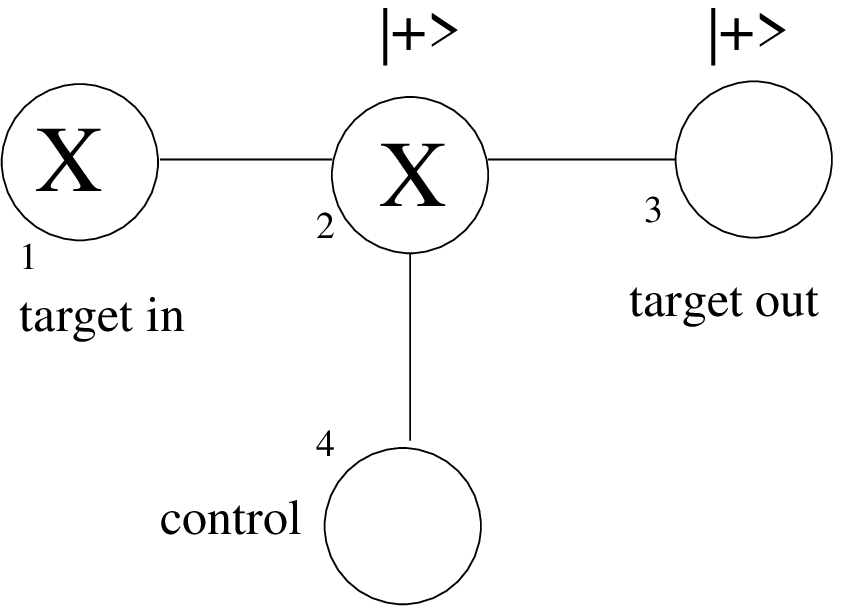}
\caption{Cluster model of \cnot gate; see: {\it Cluster.nb} and~\cite{Cluster}.}

\protect\label{ccnot}
\end{figure}

Another example of using \QDENS for cluster model studies is the
simplest \cnot gate, whose cluster model implementation from Ref.~\cite{Cluster}
is shown in Fig.~\ref{ccnot}.

This example is worked out in detail in {\it Cluster.nb}.

\section{CONCLUSION}
\label{sec11}

This simulation affords opportunities for many applications and extensions.
The basic operations and manipulations are formulated in a modular manner and,
as the illustrations and tutorial demonstrate, one can formulate and answer
many important QC questions. Application to dynamical theories based on master
equations, including environmental effects, are one challenge. Invoking and testing
measures for entanglement and probing the role of noisy entanglement, imperfect
gates, and of error correction protocols are other potential applications. Extending
the study to general types of measurements and to cluster model cases is also of
considerable interest. Finally, the description of real experimental situations by
suitable Hamiltonians and studying the stability of QC algorithms could be an
important role for future study using \QDENS.

\begin{figure}[t]
\fbox{\parbox{.8\textwidth}{\includegraphics[width=10cm]{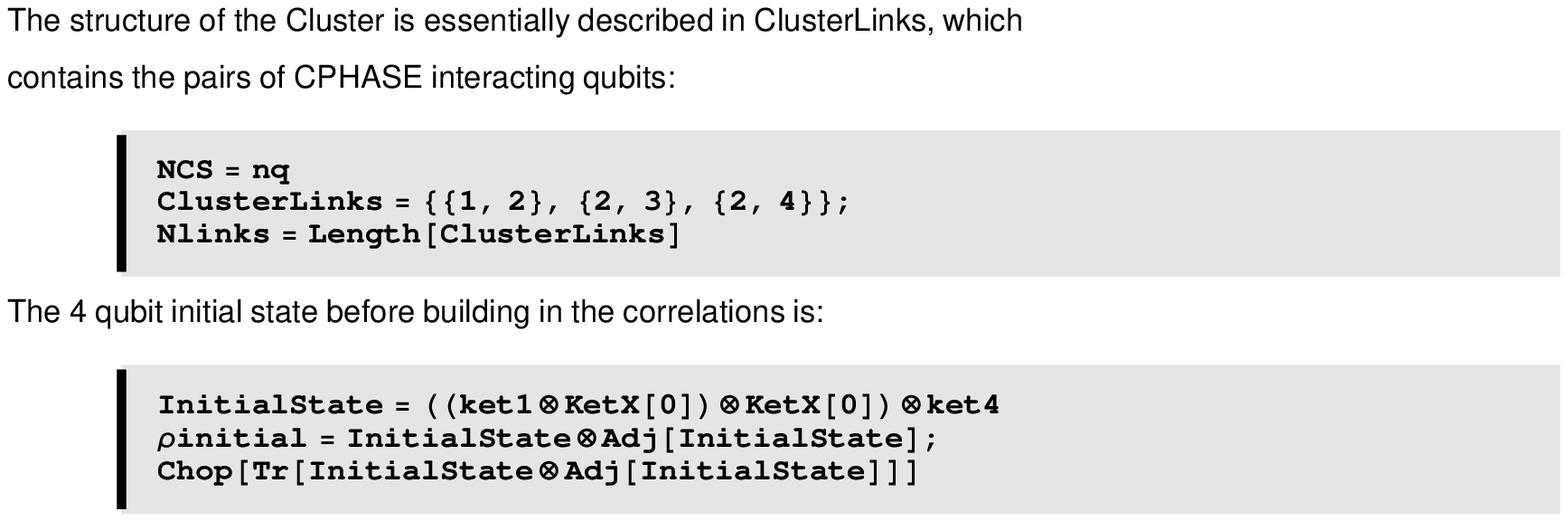}}}
\caption{Example from {\it Cluster.nb} showing  preparation of the
cluster state.}
\protect\label{cluster}
\end{figure}

\newpage
\appendix
\section{The \QDENS palette}
\begin{figure}[t]
\includegraphics[width=8cm]{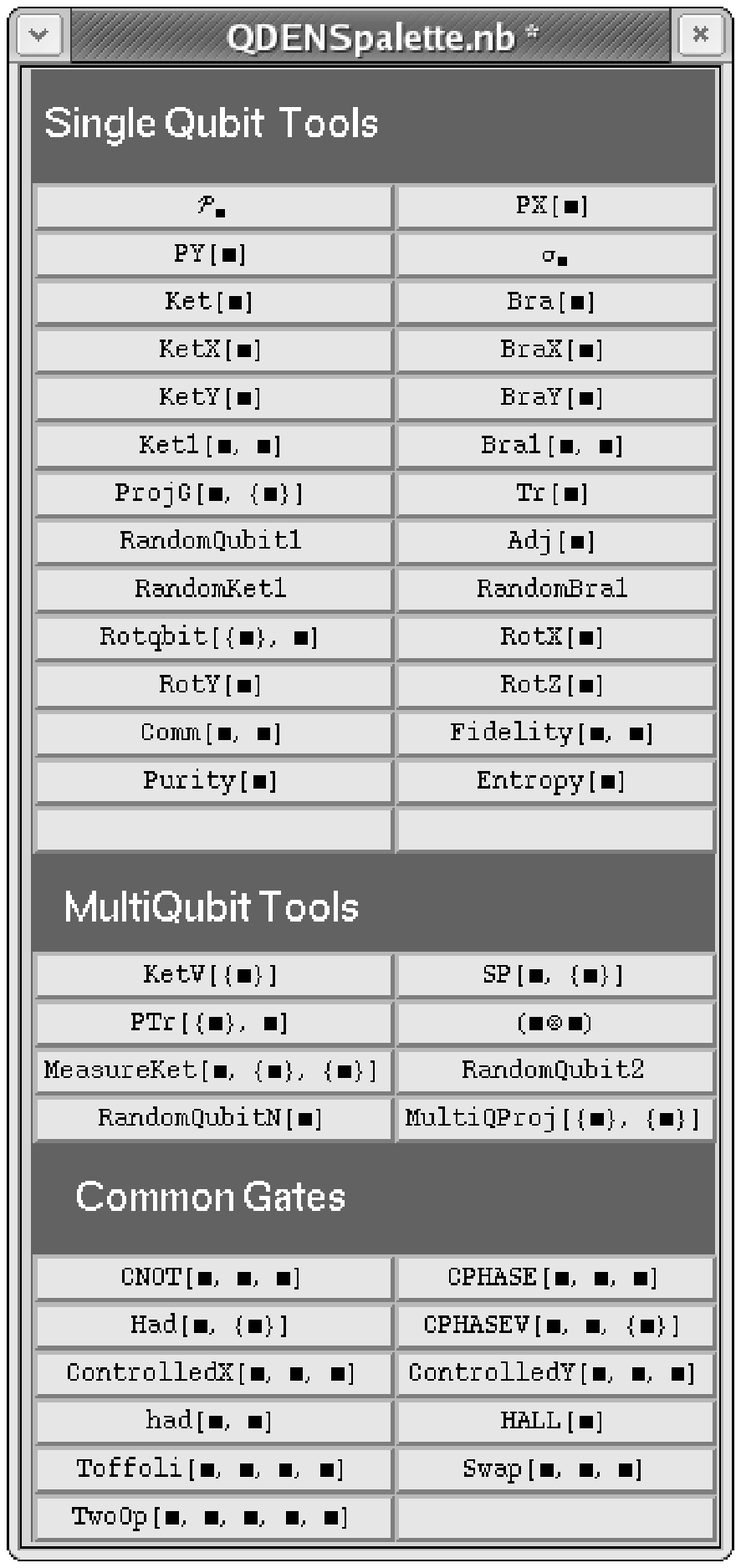}
 \caption{The \QDENS palette with the most useful commands}
\protect\label{fig:palette}
\end{figure}

The package comes together with a Mathematica Palette containing
most of the commands in the package and also some symbols that
may be useful. A snapshot of the palette is in Fig.~\ref{fig:palette}.

\section*{Acknowledgments}
 We acknowledge the participation of Matt McHugh and Mahesh Bandi at an early stage of
this project. Reza Yoosoofmiya and Sushmita Biswas made many helpful suggestions.
Insightful comments by Profs. E. Gerjuoy (PITT) and Robert B. Griffiths (CMU)
are very much appreciated.  Circuit graphs were prepared using codes from
Ref.~\cite{drawings} which we appreciate. This project was supported by the U.S. National
Science Foundation.

\end{document}